\documentclass{article}
\usepackage{arxiv}
\usepackage[utf8]{inputenc}
\usepackage[T1]{fontenc}
\usepackage{amsmath,amsfonts,amssymb}
\usepackage{graphicx}
\usepackage{booktabs}
\usepackage[table]{xcolor}
\usepackage[round]{natbib}
\usepackage{hyperref}
\hypersetup{colorlinks=true,linkcolor=blue,citecolor=blue,urlcolor=blue}
\usepackage{url}
\usepackage{microtype}
\usepackage{caption}
\let\cite\citep
\newcommand{\Description}[1]{}
\newcommand{\irbnumber}{10770}

\usepackage{multirow}
\usepackage{tikz}
\usetikzlibrary{calc}
\makeatletter
\newcommand*{\addFileDependency}[1]{\typeout{(#1)}\@addtofilelist{#1}\IfFileExists{#1}{}{\typeout{No file #1.}}}
\makeatother

\newcommand{\dg}{\ensuremath{^{\circ}}}
\newcommand{\UE}[2]{U#1$|$E#2}
\newcommand{\pholm}{p_{\mathrm{Holm}}}

\definecolor{qp}{HTML}{0072B2}
\definecolor{fv}{HTML}{D55E00}
\definecolor{pe}{HTML}{009E73}

\begin{document}

\title{Eye Tracking Signal Quality Evaluation: Meta Quest Pro vs HTC VIVE Pro Eye vs HTC VIVE Focus Vision}
\renewcommand{\shorttitle}{Eye Tracking Signal Quality Evaluation}
\date{October 2026}
\author{
  Aleksandr Chernysh \\
  Texas State University\\
  San Marcos, TX, USA \\
  \texttt{wll46@txstate.edu}
  \And
  Mehedi Hasan Raju \\
  Texas State University\\
  San Marcos, TX, USA \\
  \texttt{m.raju@txstate.edu}
  \And
  Kamrul Hasan \\
  Texas State University\\
  San Marcos, TX, USA \\
  \texttt{kamrul.hasan@txstate.edu}
  \AND
  Md Arif Shahriar \\
  Texas State University\\
  San Marcos, TX, USA \\
  \texttt{arif.shahriar@txstate.edu}
  \And
  Proloy Karmakar \\
  Texas State University\\
  San Marcos, TX, USA \\
  \texttt{proloy@txstate.edu}
  \And
  Isayas B. Adhanom \\
  Texas State University\\
  San Marcos, TX, USA \\
  \texttt{isayas@txstate.edu}
  \AND
  Oleg Komogortsev \\
  Texas State University\\
  San Marcos, TX, USA \\
  \texttt{ok@txstate.edu}
}
\maketitle

\begin{abstract}
Gaze research in virtual reality needs eye trackers of known error. 
Thirty participants completed the same protocol on the Meta Quest Pro (QP), HTC VIVE Pro Eye (PE) and HTC VIVE Focus Vision (FV); to our knowledge, this is the first evaluation of the FV eye tracker.
In a random-saccade task, median participant mean error was 1.04\dg{} on QP, 1.50\dg{} on PE and 2.41\dg{} on FV, which remained least accurate after linear recalibration.
In a later block with a darker background, PE error rose by 0.70\dg{}.
FV samples were positively correlated: dispersion was 9.8 times the sample-to-sample root mean square.
The device clocks showed losses that mean error hides. 
QP paused for up to 6.04\,s, and FV skipped single samples. 
During instructed eye closures, QP flagged every sample valid while gaze moved 9.74\dg{}. 
Evaluations should report repeated timestamps, loss composition on the device clock and validity during eye closures.

\end{abstract}


\section{Introduction}

Eye trackers are now built into several consumer virtual reality (VR) headsets. Gaze is used for interaction \cite{raju2025Interaction}, foveated rendering, social avatars and research on attention and visual behavior \cite{adhanom2023}. The quality of the gaze signal limits these uses: a gaze-selection target has to be larger than the gaze error of most users \cite{feit2017}. Manufacturers' accuracy specifications are not always reached in practice \cite{holmqvist2012,lohr2019}, so the signal has to be measured.

Evaluations of single headsets exist for an SMI-equipped HTC Vive \cite{lohr2019}, the HTC VIVE Pro Eye (PE) \cite{schuetz2022,sipatchin2021}, the Meta Quest Pro (QP) \cite{aziz2024etra,wei2023} and the HoloLens~2 \cite{aziz2022hololens}. They differ in protocol, processing and participants. Direct comparisons cover four headsets with five participants \cite{hou2024} or the Quest Pro against the add-on eye tracker of the VIVE Focus~3 \cite{rocchi2025}. We found no published evaluation of the eye tracker integrated in the HTC VIVE Focus Vision (FV) \cite{htcfocusvision}; the search is described in Suppl.~S1.

Published values also age with the software. On hardware equivalent to the Quest Pro, \citet{lohr2026gaze} compared two generations of a Meta gaze-estimation pipeline that estimates gaze from eye images at 72\,Hz. The newer pipeline reached 0.79\dg{} accuracy and 0.20\dg{} sample-to-sample RMS, against 1.07\dg{} and 0.32\dg{} for the older one (median of participant medians). Quest Pro values measured earlier with the same protocol \cite{aziz2024etra} may therefore not describe the headset today. The Pro Eye, whose eye tracking is powered by Tobii \cite{htcproeye}, is a frequent laboratory headset \cite{schuetz2022,sipatchin2021,stein2021} and gives a second reference.

Accuracy and precision summaries also assume that every recorded row is a new measurement of the eye and that the validity flag follows the state of the eye. A mean error cannot test either assumption; the device clock of each tracker can. We ask three research questions:
\begin{description}
\item[RQ1] How do accuracy and precision of the QP, FV and PE compare when the same participants complete the same protocol, and how does error change across the field of view?
\item[RQ2] What do the device clocks show about data availability and about repeated, missing and invalid samples, including samples recorded while the eyes are closed?
\item[RQ3] How does accuracy change in a second random-saccade block, which had a darker background and came later in the session, and with raised eyebrows?
\end{description}
Our contributions are:
\begin{enumerate}
\item[C1] A paired comparison of three headsets with 30 participants, the same protocol and the same analysis. It includes the first evaluation of the FV eye tracker that we could find and covers accuracy across the field of view and precision (Sections~\ref{sec:accuracy}--\ref{sec:linearity} and~\ref{sec:precision}).
\item[C2] Diagnostics of each stream on its device clock. Each headset's clock exposes a different kind of loss that the mean error hides: stale QP rows followed by rows with timestamp zero, QP interruptions longer than 0.1\,s, FV gaps of one skipped pupil sample, and QP samples flagged valid during instructed eye closures (Sections~\ref{sec:loss} and~\ref{sec:facial}). From these results we propose three checks for evaluations of headset eye trackers (Section~\ref{sec:discussion}).
\end{enumerate}

\section{Related Work}
\label{sec:related}

\paragraph{Signal quality} Spatial accuracy is the distance between the true and the measured gaze position during a fixation. Spatial precision is the reproducibility of the measured position, and data loss is the share of missing or invalid samples \cite{holmqvist2012}. \citet{lohr2019} evaluated an SMI eye tracker in an HTC Vive with a random-saccade task and added linearity, crosstalk and a user-supplied linear recalibration. The ratio of dispersion (STD) to sample-to-sample displacement (RMS) and the shape of the spectrum describe the temporal structure of a fixation signal; both change when the signal is filtered \cite{niehorster2020noise}. Accuracy and precision depend on the participant \cite{nystrom2013,blignaut2014,hooge2019}, and \citet{feit2017} derived target sizes that work for most users. \citet{aziz2024etra} evaluated the QP and introduced user and error percentiles (\UE{}{}). For each participant the E-th percentile of the errors is computed, and the U-th percentile is then taken over participants. We use this published definition; an earlier preprint \cite{aziz2024arxiv} selects one participant by rank instead (Suppl.~S3.1).

\paragraph{Eye tracking in VR headsets} For the PE, \citet{schuetz2022} found that accuracy and precision worsened with eccentricity \citep[see also][]{sipatchin2021}. \citet{wei2023} reported a preliminary evaluation of the QP, \citet{stein2021} compared eye-tracking latency across headsets, and \citet{hou2024} compared four headsets, including QP and PE, with five participants. \citet{rocchi2025} compared the QP with the add-on eye tracker of the VIVE Focus~3 in 30 participants. They found the QP more accurate and the Focus~3 tracker more precise. Our study differs in the headsets (FV instead of the Focus~3 add-on, plus PE), in the paired three-headset design, in the protocol \cite{aziz2024etra} and in the stream diagnostics. The FV is a standalone headset with integrated eye tracking and a nominal gaze output of 120\,Hz \cite{htcfocusvision}.

\section{Methods}

\subsection{Participants}
Thirty university students took part (age $24.5 \pm 6.2$ years, range 18--47; 26 men, 4 women; 8 wore glasses and 2 contact lenses; 8 had taken part in an eye-tracking study before). Individuals with known adverse reactions to VR devices or a history of seizures were excluded. All participants gave written informed consent and received course credit, \$20 or no compensation, as they preferred. The protocol was approved by the Texas State University Institutional Review Board (protocol \irbnumber).

\subsection{Apparatus}
\label{sec:apparatus}

We used the eye trackers built into the standard versions of the three headsets (Table~\ref{tab:apparatus}). Each headset ran a custom acquisition application built in Unity 2021.3.17f1; the vendor interfaces and their installed versions are listed in Table~\ref{tab:apparatus}. The three applications present the same scene, tasks and stimuli and write the same per-sample log: per-eye gaze rays, vendor validity flags, head pose and target position. Each application reads gaze through a different vendor interface, so the stream we analyze is the joint output of the device, its runtime and our logger. A chin rest supported the head.

\begin{table}[htbp]
\caption{Eye-tracking hardware and software of the three headsets. Timestamp cadence: reciprocal of the median interval between distinct device timestamps; on PE, whose clock counts whole milliseconds (intervals of 8 or 9\,ms, median 8\,ms), sample-sequence increments per second. Effective rate: unique usable samples per second of application time in RAN127 (medians over 30 recordings).}
\label{tab:apparatus}
\small
\begin{tabular}{@{}lp{0.25\textwidth}p{0.25\textwidth}p{0.25\textwidth}@{}}
\toprule
 & Meta Quest Pro (QP) & HTC VIVE Focus Vision (FV) & HTC VIVE Pro Eye (PE) \\
\midrule
Platform & Standalone (Android) & Standalone (Android) & PC-tethered (SteamVR) \\
Eye-tracking interface & OVRPlugin (C\# wrapper 1.79.0) & VIVE OpenXR 2.5.1 eye-tracker extension & SRanipal SDK 1.3.2.0 (runtime 1.3.6.12) \\
Device clock used & Gaze-sample time & Pupil-sample time & Tracker time (1\,ms resolution) \\
Nominal rate & 90\,Hz & 120\,Hz & 120\,Hz \\
Timestamp cadence & 90\,Hz & 122\,Hz & 120\,Hz \\
Effective rate & 89\,Hz & 114\,Hz & 119\,Hz \\
\bottomrule
\end{tabular}
\end{table}

\paragraph{Gaze signals} All three applications compute the cyclopean ray. Its origin is the mean of the two eye origins, its direction the normalized sum of the two eye directions, and a cyclopean sample is valid only if both eyes are valid. On the QP the left and right rays differ by a nearly constant vergence angle of 3.06\dg{} in every participant (Suppl.~S10), so they are not two independent measurements. The FV application polls the eye tracker twice per rendered frame and writes a row only when the pupil timestamp has changed. The resulting cadence of 122\,Hz is a cadence of pupil timestamps; we could not verify it as a rate of new gaze estimates (Suppl.~S2).

\paragraph{Target geometry} Random-saccade targets were head-locked on a sphere of radius 1\,m around the head origin (the center-eye camera). The cyclopean eye origin reported by each SDK lies at a different depth relative to the head origin (Suppl.~S1). Angular error, offsets and recalibration use target angles from each device's eye origin.

\subsection{Tasks and Procedure}

\begin{figure}[htbp]
\centering
\includegraphics{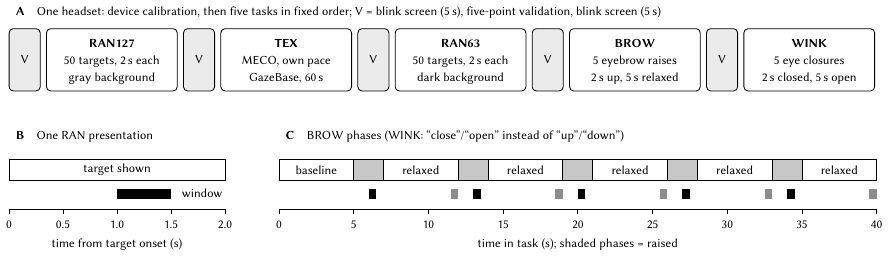}
\caption{Protocol. (A) Tasks on each headset; every participant used the three headsets in one of six counterbalanced orders. (B) Analysis window of a random-saccade (RAN) presentation. (C) Eyebrow-raising task with the windows after each ``up'' cue (black) and at the end of each relaxed phase (gray).}
\Description{Three-part schematic. Part A shows five task boxes in a row, RAN127, TEX, RAN63, BROW and WINK, each preceded by a small box for two blink screens around the validation. Part B shows a two-second timeline of one target presentation with the analysis window marked from 1.0 to 1.5 seconds. Part C shows a forty-second timeline with a five-second baseline followed by five cycles of a two-second raised phase and a five-second relaxed phase, with analysis windows marked.}
\label{fig:protocol}
\end{figure}

Participants completed five tasks on each headset in a fixed order (Figure~\ref{fig:protocol}). Before every task, the text ``PLEASE BLINK'' was shown for five seconds, followed by a validation with five targets, at the center and at $(\pm10\dg, \pm10\dg)$, and a second five-second ``PLEASE BLINK'' screen.

\paragraph{Random saccades (RAN127, RAN63)} As in \citet{aziz2024etra}, a black sphere jumped between the 49 positions of an elliptical grid spanning $\pm25\dg$ horizontally and $\pm20\dg$ vertically. The grid has a center position and three concentric elliptical rings (inner, middle, outer) with 8, 16 and 24 positions. Each target was shown for 2\,s and shrank from 2.29\dg{} to 0.57\dg{} during the first second. The center target was shown first and last, so each recording has 50 presentations; the other positions appeared in random order. In RAN127 the background was gray (127 on a 0--255 scale); RAN63 was identical except for a darker background (63). RAN63 always followed RAN127 and the reading task. On all three headsets the validation before RAN63 was shown on the dark background. The two blocks therefore differ in background and in time in session.

\paragraph{Reading (TEX)} Participants first read an English passage of the MECO corpus \cite{siegelman2022} at their own pace and ended it with the controller trigger. Each headset had its own passage (texts EN-06 on QP, EN-08 on FV and EN-12 on PE). A GazeBase passage \cite{griffith2021} followed; it was shown for 60\,s and reread if finished early. We use TEX here only for data loss.

\paragraph{Facial actions (BROW, WINK)} The eyebrow task follows \citet{aziz2024arxiv}, with a small world-locked target at the center at 1\,m. A recorded voice asked participants to raise their eyebrows (``up'') and to relax them (``down''). A five-second baseline was followed by five cycles of a two-second raise and a five-second relaxed phase. The eye-closure task had the same structure: on ``close'' participants closed their eyes and kept them closed until ``open'', two seconds later.

\paragraph{Procedure} Each of the six headset orders was used by five participants. For each headset, the experimenter adjusted the fit, ran the built-in eye-tracking calibration and placed the participant in the chin rest. One participant was recalibrated on the FV between the two random-saccade blocks; without this participant the block comparison is in Suppl.~Table~S24. Before the first task on a headset, the validation acted as a gate: mean error at most 3\dg{} with at least 80\,\% valid samples per target. The application allowed at most three gate attempts per run. In practice, after one or two failed attempts the experimenter either repeated the validation or recalibrated the tracker and restarted the run; recalibrations were not logged. The gate failed in 0 of 40 attempts on QP, 6 of 37 on FV (3 sessions) and 9 of 40 on PE (6 sessions); no run reached the three-attempt limit, and four FV runs in three sessions were restarted after one or two failures (Suppl.~Table~S5). All 450 recordings were complete; when a task had been repeated (5 recordings), we analyzed the latest complete attempt.

\subsection{Processing and Measures}

\paragraph{Usable samples and windows} We used the device clock of each eye tracker (Table~\ref{tab:apparatus}). A sample was usable if both eyes and the head pose were flagged valid and the gaze geometry was finite and non-zero. On the QP, its timestamp also had to be positive and seen for the first time in the file. QP rows with a repeated timestamp repeat the previous eye state and were removed before any metric was computed (2.3\,\% of QP rows). For RAN127 and RAN63 we analyzed the window from 1.0 to 1.5\,s after target onset, as in \citet{aziz2024etra}. The windows follow the target schedule; we did not detect fixations. Of the 9,000 scheduled windows, 8,970 contained usable samples (24, 2 and 4 missing on QP, FV and PE). Target eccentricity (head origin) is 6.7--8.3\dg, 13.3--16.7\dg{} and 20.0--25.0\dg{} on the inner, middle and outer ring.

\paragraph{Spatial accuracy} For a usable sample $i$ with cyclopean gaze direction $\mathbf{g}_i$, eye origin $\mathbf{o}_i$ and target position $\mathbf{t}$, the angular error and the error of a window $w$ with $n_w$ usable samples are
\begin{equation}
\theta_i=\arccos\frac{\mathbf{g}_i\cdot(\mathbf{t}-\mathbf{o}_i)}{\lVert\mathbf{g}_i\rVert\,\lVert\mathbf{t}-\mathbf{o}_i\rVert},
\qquad A_w=\frac{1}{n_w}\sum_{i\in w}\theta_i .
\label{eq:acc}
\end{equation}
The primary measure is the participant mean error, the mean of $A_w$ over the non-missing windows of a recording, summarized across participants by the median. The mean uses every window, including 44 windows (0.5\,\% of all RAN windows) in which a neighbor-based rule flagged gaze away from the target. The median across participants limits the influence of single recordings, such as one QP RAN63 recording with a whole-recording offset (mean error 9.95\dg). Removing the 44 windows changed the medians by at most 0.10\dg{} (Suppl.~S3.2). U$|$E is $Q_U(\{e_p(E)\}_{p=1}^{N})$, where $e_p(E)$ is the E-th percentile of the window errors of participant $p$ and $Q_U$ the U-th percentile across the $N$ participants, both with linear interpolation \cite{aziz2024etra}. Its confidence intervals come from resampling participants (10,000 resamples).

\paragraph{Precision and temporal structure} Sample-to-sample precision is the root mean square of the angles between the $m$ pairs of successive usable samples in a window. Dispersion is the angular root mean square around the normalized mean direction $\bar{\mathbf{g}}$:
\begin{equation}
\mathrm{RMS}_w=\sqrt{\frac{1}{m}\sum_{j=1}^{m}\angle(\mathbf{g}_j,\mathbf{g}_{j+1})^2},
\qquad
\mathrm{STD}_w=\sqrt{\frac{1}{n_w}\sum_{i\in w}\angle(\mathbf{g}_i,\bar{\mathbf{g}})^2}.
\label{eq:prec}
\end{equation}
For independent (white) noise $\mathrm{STD}/\mathrm{RMS}=1/\sqrt{2}$. Larger ratios mean that successive samples are positively correlated, as after temporal filtering, or that gaze drifts within the window \cite{niehorster2020noise}. The ratio is computed per window and summarized by the participant median over both blocks. We counted pairs of successive samples with identical gaze directions, which would mark held values. Following \citet{lohr2019}, we computed magnitude spectra of the three most stable one-second segments per participant and headset in RAN127. Each spectrum is summarized by the ratio of the mean magnitude at 20--39\,Hz to that at 1--4\,Hz (Suppl.~S7). Also following \citet{lohr2019}, the optimal shift is the time shift of the gaze signal (0--800\,ms, 4\,ms steps) that minimizes the mean distance between gaze and the target step function over a whole task. It adds saccade latency to the delay of the system.

\paragraph{Data loss} The expected number of samples in a recording is the device-clock span times the timestamp cadence of that recording. Lost samples fall into three disjoint categories, in percent of the expected samples. \emph{Invalid} samples are unique samples that fail the validity rules, \emph{frozen} samples are missing slots in intervals that contain repeated timestamps, and \emph{dropped} samples are the remaining gaps in the recorded stream. We also divided each recording into 250\,ms windows and counted the windows without a valid sample. Application frame durations came from the Unity time of the logged rows; a frame longer than 1.5 times the median of its recording counted as long.

\paragraph{Field of view, linearity and recalibration} For each participant we regressed window error on eccentricity (slope per 10\dg). Following \citet{lohr2019}, we regressed the 50 window centroids on target position, separately for horizontal and vertical head-local angles. We fitted their linear recalibration, $h'=a_1h+b_1v+c_1$ and $v'=a_2h+b_2v+c_2$, by least squares on every headset and tested it on held-out data. In the \emph{spatial holdout}, the positions of RAN127 alternate between two folds around each ring, and the model fitted on one fold is tested on the other. In the \emph{block transfer}, the model fitted on RAN127 is tested on RAN63.

\paragraph{Facial actions} In BROW, the raised value of a participant is the median over the five raises of the window error from 1.0 to 1.5\,s after ``up''. The relaxed value is the median over the five following relaxed phases of the window from 4.5 to 5.0\,s. In WINK, validity is the share of rows flagged valid by the device in a window after ``close''. A closure shows sustained loss of validity when it contains three consecutive invalid rows. Loss latency is the time from the cue to the first of these rows, and recovery the time from ``open'' to the first of five consecutive usable rows.

\paragraph{Statistics} All comparisons are paired (30 participants on every headset), and the participant values are skewed (Figure~\ref{fig:accuracy}), so we used rank tests instead of tests that assume normally distributed differences. Headsets were compared with a Friedman test, followed by exact Wilcoxon signed-rank tests with Holm correction within each family of three contrasts (four rings in Suppl.~Table~S22) \cite{holm1979}. We did not correct across families, which correspond to separate research questions and blocks. We treat the Holm-corrected contrasts within these families as confirmatory and all other tests as exploratory. A non-significant contrast is reported with its estimate and interval and is not read as equality. Changes within a headset (between blocks, with raised eyebrows, after recalibration) were tested with the same signed-rank test. For every paired difference we report the Hodges--Lehmann (HL) estimate \cite{hodges1963} (the median of the pairwise averages of the paired differences), the estimator that belongs to the signed-rank test, with its exact 95\,\% confidence interval from inverting the test. Confidence intervals are not adjusted for multiplicity; only the Holm-corrected p-values are. For the main contrasts we also report the rank-biserial correlation $r$ \cite{kerby2014}. Eccentricity slopes are summarized by the mean participant slope with a percentile bootstrap interval over participants (10,000 resamples) \cite{efron1986}; a random-intercept model gives the same group slopes (Suppl.~S4). The analysis used SciPy \cite{virtanen2020} and statsmodels \cite{seabold2010}. An independent reimplementation reproduced the accuracy, precision, timing, loss, linearity, recalibration and facial-action results from the raw recordings; the spectral ratios and eye-closure counts are its corrected values. It did not rerun the additional analyses A1--A12, the exact HL intervals, split-half reliability or per-eye accuracy (Suppl.~S11).

\section{Results}

\subsection{Spatial Accuracy}
\label{sec:accuracy}

\begin{table}[htbp]
\caption{Spatial accuracy (deg) per headset and block ($n = 30$).
Mean: median of participant mean errors with 95\,\% bootstrap CI.
U$|$E columns: user and error percentiles of window errors (Suppl.~S3.1).
No ring: median participant mean without the outer ring.
Without one QP recording with a whole-recording offset, RAN63 QP \UE{95}{95} is 9.23\dg.}
\label{tab:accuracy}
\small
\begin{tabular}{lcccccc}
\toprule
Block & Headset & Mean [95\,\% CI] & \UE{50}{50} & \UE{50}{95} & \UE{95}{95} & No ring \\
\midrule
\multirow{3}{*}{RAN127} & QP & 1.04 [0.86, 1.28] & 0.83 & 2.21 & 6.71  & 0.82 \\
                        & FV & 2.41 [2.11, 2.54] & 2.10 & 4.41 & 8.37  & 1.88 \\
                        & PE & 1.50 [1.29, 1.79] & 1.07 & 4.04 & 8.02  & 0.99 \\
\midrule
\multirow{3}{*}{RAN63}  & QP & 1.07 [0.86, 1.35] & 0.87 & 1.98 & 10.54 & 0.92 \\
                        & FV & 2.26 [2.07, 2.97] & 2.07 & 4.56 & 10.52 & 1.89 \\
                        & PE & 2.28 [1.87, 2.56] & 1.78 & 5.27 & 10.26 & 1.67 \\
\bottomrule
\end{tabular}
\end{table}

\begin{figure}[htbp]
\centering
\includegraphics[width=0.4\linewidth]{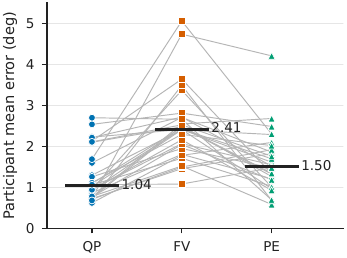}
\caption{Participant mean error (deg) in RAN127 ($n = 30$).
Gray lines connect each participant; bars mark medians.
In all figures: QP blue circles, FV orange squares, PE green triangles.
Error by target eccentricity: Suppl.~S4.}
\label{fig:accuracy}
\end{figure}

In RAN127 the median participant mean error was 1.04\dg{} on QP, 2.41\dg{} on FV and 1.50\dg{} on PE (Table~\ref{tab:accuracy}, Figure~\ref{fig:accuracy}). Error differed between headsets (Friedman $\chi^2=32.1$, $p<.001$). QP was more accurate than FV (HL $-1.09\dg$ [$-1.43$, $-0.81$], $r=-0.99$, $\pholm<.001$; lower error in 29 of 30 participants). PE was also more accurate than FV ($-0.83\dg$ [$-1.18$, $-0.51$], $r=-0.88$, $\pholm<.001$). QP minus PE was $-0.28\dg$ [$-0.59$, $+0.00$] ($r=-0.41$, $\pholm=.052$). When windows with less than 80\,\% of the nominal number of samples were left out, $\pholm$ became .047 (Suppl.~Table~S9), so we draw no conclusion about QP against PE in the first block. In RAN63 the medians were 1.07\dg, 2.26\dg{} and 2.28\dg{} (Friedman $\chi^2=20.3$, $p<.001$). QP was more accurate than FV ($-1.28\dg$ [$-1.77$, $-0.77$], $\pholm<.001$) and PE ($-0.94\dg$ [$-1.36$, $-0.50$], $\pholm<.001$), also without the outer ring (QP minus PE $-0.67\dg$ [$-1.00$, $-0.29$], $\pholm=.006$). PE minus FV was $-0.37\dg$ [$-0.89$, $+0.10$] ($\pholm=.124$). Participant ranks agreed weakly across headsets (Kendall's $W=0.47$, $p=.075$). For RAN127 error centered on the median of its headset, the Friedman test across the first, second and third position in the session gave $\chi^2=4.20$ ($p=.122$); the largest contrast, second minus first position, was $-0.27\dg$ (unadjusted interval [$-0.63$, $-0.04$]; $\pholm=.121$; Suppl.~Table~S4).

The percentiles describe the tails. In RAN127 the median across participants of each participant's 95th-percentile window error (\UE{50}{95}) was 2.21\dg{} (QP), 4.41\dg{} (FV) and 4.04\dg{} (PE). Covering 95\,\% of the windows of 95\,\% of the participants (\UE{95}{95}) required 6.71\dg{} [5.05, 7.56], 8.37\dg{} [5.92, 14.53] and 8.02\dg{} [5.24, 12.48]. These are 8.1, 4.0 and 7.5 times the \UE{50}{50}; with 30 participants the upper percentiles are uncertain.

\subsection{Accuracy Across the Field of View}
\label{sec:fov}

\begin{figure}[htbp]
\centering
\includegraphics[width=\textwidth]{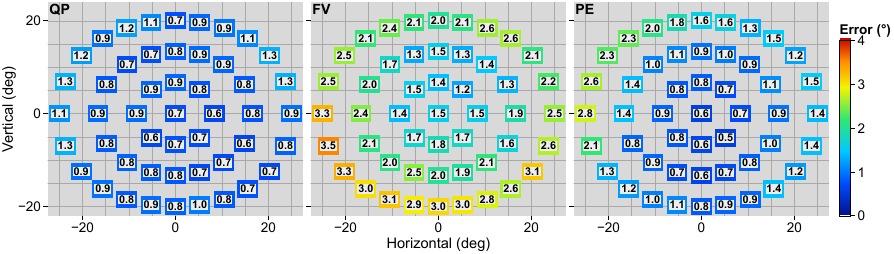}
\caption{Accuracy at each of the 49 target positions in RAN127: median over 30 participants of the participant mean error (deg) at that position. Axes give target angles at the head origin. RAN63 maps and maps of single participants at ranks 15, 23 and 29 are in Suppl.~S4; offset vectors are in Suppl.~S5.}
\label{fig:maps}
\end{figure}

Error increased from the center to the periphery on all headsets (Figure~\ref{fig:maps}). In RAN127 the median participant error was 0.71\dg{} at the center and 1.18\dg{} on the outer ring for QP, 1.52\dg{} and 2.86\dg{} for FV, and 0.55\dg{} and 2.05\dg{} for PE. The position medians ranged from 0.63\dg{} to 1.31\dg{} on QP, from 1.23\dg{} to 3.50\dg{} on FV and from 0.53\dg{} to 2.75\dg{} on PE. The mean participant slope was 0.35\dg{} per 10\dg{} of eccentricity on QP [0.23, 0.48], 0.72\dg{} on FV [0.52, 0.92] and 0.81\dg{} on PE [0.65, 0.98]. Slopes were steeper on FV than on QP (HL 0.39\dg{} per 10\dg{} [0.18, 0.60], $\pholm=.004$) and on PE than on QP (0.44\dg{} [0.21, 0.68], $\pholm=.001$). PE minus FV was 0.10\dg{} [$-0.20$, 0.37] ($\pholm=.529$). Without the outer ring (eccentricity of 20\dg{} and more), the medians fell to 0.82\dg, 1.88\dg{} and 0.99\dg. QP minus PE was then $-0.06\dg$ [$-0.27$, 0.22] ($\pholm=.670$), and FV remained less accurate than both ($\pholm<.001$).

As an orientation for target sizes, the radius that contained 95\,\% of all RAN127 window errors was 1.49\dg{} at the center and 5.73\dg{} on the outer ring for PE. It was 2.27\dg{} and 4.71\dg{} for QP and 3.37\dg{} and 6.11\dg{} for FV (all rings: Suppl.~S4). These radii pool windows across participants. An interface that should work for nearly all users needs the \UE{95}{95} instead (6.71--8.37\dg{} in RAN127 and 10.26--10.54\dg{} in RAN63, Table~\ref{tab:accuracy}). Our tasks contained no gaze selection, so target sizes derived from either measure still have to be confirmed in an interaction test.

\subsection{Systematic Error and Recalibration}
\label{sec:linearity}

Gaze was nearly linear in target position on all headsets (median adjusted $R^2$ of at least 0.987 in RAN127; Suppl.~Table~S17). FV compressed gaze toward the center. Its mean vertical slope was 0.903 with target angles from the device eye origin and 0.923 from the head origin, against 0.979 and 0.979 for QP and 1.014 and 0.975 for PE. The vertical slopes differed between headsets under both conventions (Friedman $\chi^2=30.2$, $p<.001$, and $\chi^2=13.1$, $p=.001$). The horizontal contrast depended on the convention (eye origin $\chi^2=20.9$, $p<.001$; head origin $\chi^2=2.4$, $p=.301$; Suppl.~S5). The median length of the offset vector over positions was 0.46\dg{} on QP, 1.56\dg{} on FV and 0.40\dg{} on PE. QP placed gaze below the target (median vertical offset $-0.41\dg$, below in 25 of 30 participants).

A linear recalibration fitted on one fold of positions and tested on the other lowered the participant mean error on all three headsets. The change was $-0.82\dg$ [$-1.06$, $-0.65$] on FV (median from 2.41\dg{} to 1.48\dg; no participant worse), $-0.29\dg$ [$-0.41$, $-0.17$] on PE (5 participants worse) and $-0.15\dg$ [$-0.23$, $-0.07$] on QP (8 worse; all $\pholm\leq.003$). Any linear fit is expected to gain most where the error is closest to linear, as on FV. Fitted on RAN127 and applied to RAN63, the correction changed the error by $-0.41\dg$ [$-0.64$, $-0.25$] on FV and $-0.46\dg$ [$-0.63$, $-0.30$] on PE (3 participants worse on each). On QP the change was $-0.04\dg$ [$-0.10$, $+0.02$] ($\pholm=.191$; 11 worse). We did not test a calibration with few targets, as an application would use.

\subsection{Change in the Second Block}
\label{sec:block}

From RAN127 to RAN63 the participant mean error changed by HL $+0.10\dg$ [$-0.06$, $+0.27$] on QP ($\pholm=.360$) and $+0.14\dg$ [$-0.04$, $+0.38$] on FV ($\pholm=.219$). On PE it increased by $+0.70\dg$ [$+0.47$, $+0.98$] ($r=0.86$, $\pholm<.001$; in 26 of 30 participants). The PE increase exceeded the QP change by 0.56\dg{} [0.31, 0.86] ($\pholm=.002$) and the FV change by 0.48\dg{} [0.12, 0.88] ($\pholm=.024$). The second contrast became $\pholm=.047$ and .059 under two rules that leave out windows with less than 80\,\% of the expected samples (Suppl.~Table~S10). The FV median fell (2.41\dg{} to 2.26\dg) while the FV HL was positive: the HL of paired changes need not equal the difference of medians. On PE the increase appeared on every ring, from HL $+0.49\dg$ [0.22, 0.84] at the center to $+0.76\dg$ [0.52, 1.06] on the outer ring (Suppl.~S6), and sample-to-sample RMS rose from 0.17\dg{} to 0.23\dg{} ($p<.001$). Between the first and second half of the presentations within a block, the largest change was on PE in RAN127 (HL $+0.15\dg$ [0.01, 0.30], $\pholm=.109$; all others $\pholm\geq.926$; Suppl.~Table~S23).

Pupil diameter grew in the second block on both HTC headsets; QP logs no pupil size. On PE it grew from 3.61 to 4.71\,mm (30 of 30 participants; exact Wilcoxon $p<.001$). On FV the values grew from 3.86 to 5.17 (29 of 30; $p<.001$). The FV interface reports millimeters \cite{htcpupil}, but its scale was not validated, so we compare FV values only within the headset. The change in pupil size did not correlate with the change in error across participants ($\rho=0.28$, $p=.140$ on PE; $\rho=-0.14$, $p=.449$ on FV). The PE validation error did not change before TEX (HL $+0.00\dg$ [$-0.09$, 0.09], $\pholm=.984$), rose before RAN63, its only validation on the dark background ($+0.35\dg$ [0.14, 0.55], $\pholm=.003$), and fell again before BROW, on gray ($-0.23\dg$ [$-0.40$, $-0.09$], $\pholm=.009$; Suppl.~Table~S27). The PE error thus did not grow steadily with time; the cause of the increase remains open (Section~\ref{sec:limitations}).

\subsection{Precision and Temporal Structure}
\label{sec:precision}

\begin{figure}[htbp]
\centering
\includegraphics[width=0.75\textwidth]{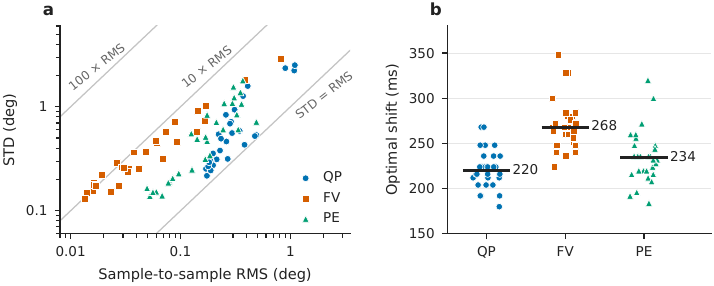}
\caption{Temporal structure in RAN127 ($n = 30$ per headset).
(a) Dispersion (STD) against sample-to-sample precision (RMS), one point per participant (means over windows, deg); diagonals mark STD = 1, 10 and 100 $\times$ RMS.
(b) Optimal gaze--target shift per participant (ms; saccade latency plus system delay); bars and numbers mark medians.
Spectra: Suppl.~S7.}
\label{fig:temporal}
\end{figure}

The median participant RMS in RAN127 was 0.26\dg{} on QP, 0.04\dg{} on FV and 0.17\dg{} on PE, and the median STD was 0.48\dg, 0.28\dg{} and 0.50\dg{} (Figure~\ref{fig:temporal}a). FV had the lowest RMS, whereas its STD was only 0.20\dg{} below that of QP. The median STD/RMS ratio over both blocks was 9.8 [interquartile range 8.5--10.4] on FV, against 1.3 [1.2--1.6] on QP and 2.4 [2.2--2.7] on PE. In the spectra of stable segments the FV magnitude fell steeply above a few hertz. The ratio of mean magnitude at 20--39\,Hz to that at 1--4\,Hz was 0.021 on FV, 0.281 on QP and 0.093 on PE. The FV ratio was below the QP ratio in 30 and the PE ratio in 28 of 30 participants (Suppl.~S7). The median optimal shift in RAN127 was 268\,ms on FV [IQR 257--279], 220\,ms on QP [212--233] and 234\,ms on PE [217--247] (Figure~\ref{fig:temporal}b; RAN63: 276, 224 and 236\,ms; descriptive).

Held values do not explain the low FV RMS. A median of 1.8 per 10,000 successive FV sample pairs [IQR 1.8--4.9] had identical gaze directions (QP: none; PE: 7 of 170,513 pairs), and no run of identical values was longer than 3 samples. RMS computed on the remaining pairs was unchanged (0.04\dg; Suppl.~S7). FV increments were positively correlated (median lag-1 autocorrelation $+0.29$ in 30 of 30 participants; QP $-0.08$, PE $-0.06$), and only 1.8\,\% of successive increments were equal, against about half under linear interpolation of a half-rate stream (Suppl.~Table~S30; descriptive). These results, the STD/RMS ratio and the spectrum are consistent with temporal filtering or non-linear interpolation of a lower-rate stream, not with sample-and-hold or linear interpolation. We did not test a filter directly.

\subsection{Data Loss and the Device Clock}
\label{sec:loss}

Data loss differed in composition (Table~\ref{tab:loss}): frozen samples on QP, dropped on FV and invalid on PE.

\begin{table}[htbp]
\caption{Data loss per task (\% of expected samples on the device clock, $n = 30$).
Inv., Fr., Dr.: median invalid, frozen and dropped samples. Total: median of their sum. Max: maximum total.
``--'': zero; ``$<$0.1'': above zero and below 0.05.
Category medians need not add up to the total; small negative dropped estimates from clock resolution count as zero.
WINK includes the instructed eye closures.
For FV, the data cannot tell whether the device or the logger lost the dropped samples (Section~\ref{sec:loss}).}
\label{tab:loss}
\small
\setlength{\tabcolsep}{4.2pt}
\begin{tabular}{lrrrrrrrrrrrrrrr}
\toprule
 & \multicolumn{5}{c}{QP} & \multicolumn{5}{c}{FV} & \multicolumn{5}{c}{PE} \\
\cmidrule(lr){2-6}\cmidrule(lr){7-11}\cmidrule(lr){12-16}
Task & Inv. & Fr. & Dr. & Total & Max & Inv. & Fr. & Dr. & Total & Max & Inv. & Fr. & Dr. & Total & Max \\
\midrule
RAN127 & -- & 1.3 & $<$0.1 & 1.3 & 10.8 & --     & -- & 5.9 & 6.0  & 8.2  & 1.0  & -- & -- & 1.0  & 22.9 \\
TEX    & -- & 0.5 & 0.1    & 0.5 & 10.9 & --     & -- & 6.0 & 6.0  & 9.4  & 1.3  & -- & -- & 1.3  & 13.4 \\
RAN63  & -- & 2.7 & $<$0.1 & 2.7 & 17.7 & $<$0.1 & -- & 6.4 & 6.4  & 9.4  & 2.0  & -- & -- & 2.0  & 24.7 \\
BROW   & -- & 0.5 & $<$0.1 & 0.5 & 7.9  & 0.3    & -- & 6.6 & 7.3  & 15.8 & 1.9  & -- & -- & 1.9  & 67.7 \\
WINK   & -- & 2.8 & $<$0.1 & 2.8 & 13.4 & 23.3   & -- & 7.1 & 30.6 & 43.8 & 27.4 & -- & -- & 27.4 & 36.7 \\
\bottomrule
\end{tabular}
\end{table}

\begin{figure}[htbp]
\centering
\begin{minipage}[b]{0.58\textwidth}
\centering
{\small\textbf{a}}\par
\includegraphics[width=\linewidth]{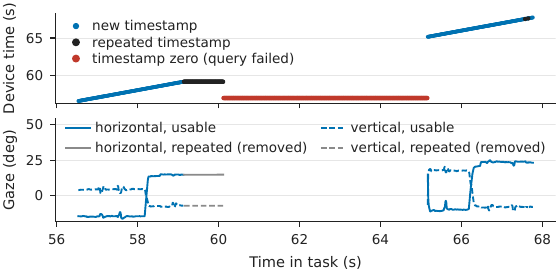}
\end{minipage}\hfill
\begin{minipage}[b]{0.29\textwidth}
\centering
{\small\textbf{b}}\par
\includegraphics[width=\linewidth]{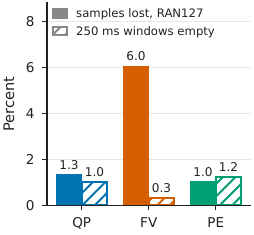}
\end{minipage}
\caption{Data loss on the device clock.
(a) Longest QP interruption (6.04\,s, one participant, RAN63). Top: the device timestamp repeats for about 1\,s, then queries return zero (drawn at the bottom) until tracking resumes. Bottom: logged gaze (deg), frozen during the repeats. All these rows are removed before analysis.
(b) Filled: median share of expected samples lost in RAN127 ($n = 30$). Hatched: mean share of 250\,ms windows without a valid sample (120 recordings, all tasks except WINK). Both in \%.}
\label{fig:interruption}
\end{figure}

Over all tasks QP had 233 intervals longer than 0.1\,s between unique device timestamps, in 26 of 30 participants; FV had one such interval and PE none. Of these, 230 lie between fresh timestamps of the task rows; the other 3 start at the first row of a task, whose timestamp had already appeared in the rows logged before the task. In 133 of the 233, the timestamp first repeated, and then the queries returned timestamp zero until tracking resumed (Figure~\ref{fig:interruption}a). These intervals lasted a median of 1.02\,s (longest 6.04\,s). Rows with timestamp zero carried no gaze geometry and no valid flag (0 of 967 rows). The other 100 intervals consisted of repeated timestamps only (median 0.19\,s). None of the 230 interruptions within task rows started in a long application frame (Suppl.~S8). Rows with a repeated timestamp carry the previous gaze unchanged. Keeping them changed the QP \UE{50}{50} in RAN127 by less than 0.01\dg{} and the \UE{95}{95} from 6.71\dg{} to 6.68\dg. In RAN63, it raised the \UE{95}{95} from 10.54\dg{} to 12.52\dg, because 13 otherwise empty windows were then filled with frozen gaze (Suppl.~S8).

Of the intervals between successive FV rows, 94.4\,\% spanned one tick of the pupil timestamp and 5.5\,\% spanned two. Long application frames were rare (2.3 per 10,000 frames; median frame 11\,ms) and accounted for 0.1\,\% of the skipped ticks in RAN127. Instead, 98.2\,\% of the two-tick intervals followed one polling pattern. The second poll of a frame, 9\,ms after the first, returned the previous sample; the first poll of the next frame, 2\,ms later, returned a sample two ticks later (Suppl.~S8). The skipped sample was therefore exposed by the interface for at most about 2\,ms, or not at all; our data cannot tell which. The 5.9\,\% dropped FV samples in RAN127 are thus gaps in the recorded stream that frame stalls do not explain. In RAN127 these 5.9\,\% consist of the two-tick intervals (5.1\,\% of intervals, Suppl.~Table~S35; 4.6\,\% of expected samples), longer gaps (below 0.1\,\%) and 1.3\,\% from the denominator: the 122\,Hz cadence is the reciprocal of the median interval, whereas one-tick intervals average 120\,Hz; at 120\,Hz the dropped share would equal the two-tick share.

PE losses were invalid samples, concentrated in a few recordings; the worst recording outside WINK, in BROW, lost 67.7\,\% of its samples. Averaged over recordings (WINK excluded), 1.0\,\% (QP), 0.3\,\% (FV) and 1.2\,\% (PE) of the 250\,ms windows contained no valid sample (Figure~\ref{fig:interruption}b). FV, with the largest median loss, had the fewest empty windows: its losses are single samples, whereas QP and PE lose consecutive samples.

\subsection{Eyebrow Raising and Eye Closure}
\label{sec:facial}

\begin{figure}[htbp]
\begin{minipage}[t]{0.40\textwidth}\centering
\vspace{0pt}
\includegraphics[width=\linewidth]{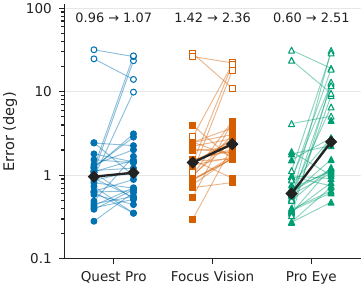}
\captionof{figure}{Eyebrow raising: error (deg, logarithmic axis) with relaxed (left) and raised (right) eyebrows for each of 30 participants per headset. Diamonds and numbers are medians; open symbols mark participants with raised values above 5\dg, who were not excluded.}
\Description{For each of the three headsets, paired points per participant connect the error with relaxed and raised eyebrows on a logarithmic axis. Medians change from 0.96 to 1.07 degrees on QP, from 1.42 to 2.36 on FV and from 0.60 to 2.51 on PE.}
\label{fig:facial}
\end{minipage}\hfill
\begin{minipage}[t]{0.57\textwidth}\centering
\vspace{0pt}
\captionof{table}{Eye closure (WINK), 30 participants with five closures each. Share of rows flagged valid after the ``close'' cue (medians over participants); closures with sustained loss of validity (three consecutive invalid rows) among closures with recorded rows; latency of the loss and of the recovery (medians, s). Last rows: QP gaze 1.0--2.0\,s after ``close'' against the last 0.5\,s before it (medians over participants of the median over five closures). Closure onset from eye openness uses another definition (Suppl.~Table~S37).}
\label{tab:wink}
\small
\setlength{\tabcolsep}{2pt}
\begin{tabular}{@{}lrrr@{}}
\toprule
 & QP & FV & PE \\
\midrule
Valid, 0.5--1.5\,s after cue (\%) & 100.0 & 41.8 & 16.6 \\
Valid, 1.0--2.0\,s after cue (\%) & 100.0 & 1.5 & 0.0 \\
Closures with sustained loss & 0 of 150 & 148 of 149 & 150 of 150 \\
Loss latency (s) & -- & 0.92 & 0.68 \\
Recovery after ``open'' (s) & -- & 0.93 & 0.76 \\
\addlinespace
QP fresh samples per second, closed & 90 &  &  \\
QP distance from pre-closure gaze (\dg) & 9.74 &  &  \\
QP STD, open / closed (\dg) & 0.23 / 2.98 &  &  \\
\bottomrule
\end{tabular}
\end{minipage}
\end{figure}

\paragraph{Eyebrow raising} With eyebrows relaxed and raised, the median error was 0.96\dg{} and 1.07\dg{} on QP, 1.42\dg{} and 2.36\dg{} on FV and 0.60\dg{} and 2.51\dg{} on PE (Figure~\ref{fig:facial}). The paired increase was HL $+0.11\dg$ [$-0.14$, 0.66] on QP ($\pholm=.404$), $+0.78\dg$ [0.30, 1.58] on FV ($\pholm=.013$) and $+1.24\dg$ [0.44, 6.12] on PE ($\pholm=.002$; 29 complete pairs). The raised error exceeded 5\dg{} in 5, 5 and 11 participants; no participant was excluded, and the exclusion rule of earlier work lowers the raised medians (Suppl.~S9), and with it the PE increase is not significant ($p=.110$, 12 pairs). The PE increase minus the FV increase was HL 0.69\dg{} [$-0.51$, 4.85] ($\pholm=.530$), and minus the QP increase 1.06\dg{} [0.18, 4.36] ($\pholm=.061$). The data therefore do not rank the headsets by eyebrow sensitivity.

\paragraph{Eye closure} From 1.0 to 2.0\,s after ``close'', the median share of rows flagged valid was 100.0\,\% on QP, 1.5\,\% on FV and 0.0\,\% on PE (Table~\ref{tab:wink}). On FV 148 of the 149 closures with recorded rows and on PE all 150 showed sustained loss of validity. The loss started a median of 0.92 and 0.68\,s after the cue, and validity returned 0.93 and 0.76\,s after ``open''. No QP closure showed sustained loss; its only invalid rows were six rows with failed queries in four closures. During the closures QP delivered fresh samples at 90 per second, with no repeated timestamps (median 0\,\%), and every sample was flagged valid. Gaze moved a median of 9.74\dg{} [IQR 7.36--12.50] away from its position before the closure, downward in 25 of 30 participants (median vertical change $-6.6\dg$). Its STD rose from 0.23\dg{} with open eyes to 2.98\dg{} (higher in 30 of 30 participants, $p<.001$). The distance did not grow over the closure (change from the first to the last 0.25\,s: median 0.12\dg, $p=.685$). QP gaze confidence stayed at 0.998 in all closed-eye rows of the median participant. The QP flag also stayed set during the two 5-s blink screens before every task (median 100.0\,\% valid rows, 300 screens; blinks not recorded; Suppl.~Table~S39).

\section{Discussion}
\label{sec:discussion}

\subsection{Main findings}
In the first block, FV was the least accurate headset (median participant mean error 2.41\dg{}, against 1.04\dg{} on QP and 1.50\dg{} on PE); QP and PE did not differ reliably, overall or in the central field. The accuracy summaries hid what the device clocks showed: QP interruptions of up to several seconds, eye closures that QP did not flag as invalid, and FV samples missing from the stream at an unknown point between device and logger.

\subsection{What the device clocks add} On QP, stale rows hardly changed the RAN127 accuracy summaries, yet the same stream contained interruptions of up to several seconds (Section~\ref{sec:loss}). The QP validity flag did not mark the instructed 2-s closures or the blink screens, and gaze during the closures had a STD 13 times higher than with open eyes (Section~\ref{sec:facial}). A blink detector or gaze filter based on the QP validity flag keeps these samples. Meta's face-tracking interface exposes eye-closure weights separately \cite{metafaceexpressions}; we did not record them. On FV, the logger sees pupil timestamps, so the skipped samples cannot be assigned to the device or the interface. We therefore recommend three checks for evaluations of headset eye trackers, next to the U$|$E summaries: repeated timestamps, loss composition on the device clock and the validity flag during instructed closures.

\subsection{FV accuracy and signal} FV gaze was compressed toward the center. A linear recalibration lowered the FV median error by 38.5\,\% in the spatial holdout (2.41\dg{} to 1.48\dg) and by 18.2\,\% when the RAN127 fit was applied to RAN63 (2.26\dg{} to 1.85\dg). The holdout change was larger than on QP or PE (descriptive), but corrected FV remained less accurate than uncorrected QP (1.04\dg; Section~\ref{sec:linearity}). The low FV RMS does not describe its dispersion (STD 0.28\dg), so FV precision should be reported with both measures. The FV optimal shift was 48\,ms longer than on QP (difference of group medians, descriptive). This shift includes saccade latency and does not isolate a signal delay. If the output is filtered, velocity-based event detection and peak saccade velocity are affected; we did not measure either.

\subsection{Comparison with prior work} \citet{aziz2024etra} report a QP \UE{50}{50} of 1.08\dg{} and a \UE{95}{95} of 17.06\dg{} for 78 participants, with the same grid and analysis window. Our values are 0.83\dg{} and 6.71\dg. Stream cleaning does not explain the difference: with repeated rows kept, our \UE{95}{95} was 6.68\dg{} (Section~\ref{sec:loss}). Both studies used a chin rest and the built-in calibration. Untested candidates for the difference are the sample (30 participants, 26 men, against 78 college-aged participants), our validation gate before the first task, the session content (three headsets in one session here) and the gaze-estimation pipeline. On hardware equivalent to the QP, a newer Meta pipeline was more accurate than an older one \citep{lohr2026gaze}; we do not know which pipeline the QP runtime used in either study. \citet{rocchi2025}, with head-free and head-constrained conditions and targets at several depths, found the QP more accurate and the Focus~3 add-on tracker more precise. We found the same order for QP against FV, but we found no source stating whether the Focus~3 add-on and the FV share tracking hardware, so our FV results do not replicate theirs. For PE, \citet{schuetz2022} report a mean binocular error of 1.10\dg{} and 0.97\dg{} in their inner 20\dg{} field. Our median of participant means is 1.50\dg{} overall and 0.99\dg{} without the outer ring, and both studies find larger error in the periphery.

\subsection{Choosing a Headset}

\begin{table}[htbp]
\caption{Measured values for common uses of gaze in VR ($n = 30$ per headset; RAN127 unless stated).
Paired changes are Hodges--Lehmann estimates; only the PE block change differed from the other headsets.}
\label{tab:summary}
\small
\begin{tabular}{lrrrl}
\toprule
Use and measure & QP & FV & PE & Section \\
\midrule
\multicolumn{5}{l}{\emph{Gaze selection and pointing}} \\
\quad Median participant mean error & 1.04\dg & 2.41\dg & 1.50\dg & \ref{sec:accuracy} \\
\quad \UE{50}{95} / \UE{95}{95} & 2.21\dg{} / 6.71\dg & 4.41\dg{} / 8.37\dg & 4.04\dg{} / 8.02\dg & \ref{sec:accuracy} \\
\quad Error increase per 10\dg{} eccentricity & 0.35\dg & 0.72\dg & 0.81\dg & \ref{sec:fov} \\
\quad 95\,\% error radius, center / outer ring & 2.27\dg{} / 4.71\dg & 3.37\dg{} / 6.11\dg & 1.49\dg{} / 5.73\dg & \ref{sec:fov} \\
\quad Change with linear recalibration & $-0.15\dg$ & $-0.82\dg$ & $-0.29\dg$ & \ref{sec:linearity} \\
\midrule
\multicolumn{5}{l}{\emph{Change between conditions}} \\
\quad Second minus first block & $+0.10\dg$ & $+0.14\dg$ & $+0.70\dg$ & \ref{sec:block} \\
\quad Raised minus relaxed eyebrows & $+0.11\dg$ & $+0.78\dg$ & $+1.24\dg$ & \ref{sec:facial} \\
\midrule
\multicolumn{5}{l}{\emph{Availability and validity}} \\
\quad 250\,ms windows without valid data (except WINK) & 1.0\,\% & 0.3\,\% & 1.2\,\% & \ref{sec:loss} \\
\quad Interruptions $>$0.1\,s (all tasks) & 233 & 1 & 0 & \ref{sec:loss} \\
\quad Closures with sustained loss of validity & 0 of 150 & 148 of 149 & 150 of 150 & \ref{sec:facial} \\
\midrule
\multicolumn{5}{l}{\emph{Eye-movement measures}} \\
\quad Sample-to-sample RMS & 0.26\dg & 0.04\dg & 0.17\dg & \ref{sec:precision} \\
\quad STD/RMS ratio, both blocks (white noise: 0.71) & 1.3 & 9.8 & 2.4 & \ref{sec:precision} \\
\quad Optimal gaze--target shift (descriptive) & 220\,ms & 268\,ms & 234\,ms & \ref{sec:precision} \\
\bottomrule
\end{tabular}
\end{table}

Table~\ref{tab:summary} collects values for three common uses (chin rest, device calibration, no gaze-selection task).  
In the central field of the first block, QP minus PE was $-0.06\dg$ ($\pholm=.670$), and both were more accurate than uncorrected FV. In the second block QP was more accurate than PE, also without the outer ring, and PE minus FV was $-0.37\dg$ ($\pholm=.124$; Section~\ref{sec:accuracy}). For nearly all users the error budget is the \UE{95}{95}: 6.71--8.37\dg{} in the first and 10.26--10.54\dg{} in the second block, and it grows toward the periphery on all three headsets. Across conditions, only the PE block change differed from the changes of the other two headsets; the eyebrow data do not rank the headsets (Section~\ref{sec:facial}). For availability, QP drops out for up to several seconds, and its validity flag did not mark instructed closures; the FV and PE flags did. For eye-movement measures, FV RMS is not directly comparable with the RMS of the other headsets, and the QP per-eye rays cannot be used for vergence.

\section{Limitations}
\label{sec:limitations}

A chin rest limited head movement and slippage; free head movement may lower accuracy \cite{hooge2023}. We tested one unit per headset, once per participant (versions in Table~\ref{tab:apparatus}). Fixed block order confounds the RAN63--RAN127 difference with time in session and the preceding background. The validation gate removed poor FV calibrations more often than others (Suppl.~Table~S5), which if anything favors FV. Display luminance was not measured. The PE increase may come from pupil-size dependence \cite{wyatt2010,drewes2014,choe2016} or block order; we could not separate them (Section~\ref{sec:block}). Without eye video or headset tracking, we cannot attribute the eyebrow effect to headset movement \cite{niehorster2020slippage} or eye-image occlusion. Spoken cues timed the eye closures.

We evaluated each interface's output as logged by our application. Testing whether the skipped FV samples ever existed would require polling the interface far more often than twice per frame. Target angles use the eye origin reported by each SDK. With head-origin angles, the median centroid error changed by less than 0.2\dg{} on every headset, and the headset order did not change. Two centroid comparisons near the threshold crossed it: QP against PE in RAN127 ($\pholm=.040$ eye-origin, .073 head-origin) and PE against FV in RAN63 (.084 and .050; Suppl.~S5).

The participants were 30 university students aged 18--47, 26 of them men, which limits generalization to other populations.

\section{Conclusion}

In the first block, FV was the least accurate of the three headsets. A linear recalibration lowered its median error by 38.5\,\% in a spatial holdout (2.41\dg{} to 1.48\dg) and by 18.2\,\% in the second block; corrected FV remained less accurate than uncorrected QP. In the second block, PE error rose by HL 0.70\dg{} [0.47, 0.98]; our design does not identify the cause. FV precision should be reported with both RMS and STD. The QP validity flag stayed set during instructed 2-s eye closures and blink screens, so a blink filter based on it keeps closed-eye samples; the FV and PE flags marked the closures. Evaluations of headset eye trackers should add three checks on the device clock to the accuracy summaries: repeated timestamps, the composition of data loss, and the validity flag during instructed eye closure.

\section*{Privacy and Ethics Statement}
Eye movements are biometric data that can identify a person and reveal attributes such as attention, fatigue or health. Knowledge of signal quality supports beneficial uses of gaze as well as such inferences. The study was approved by the Texas State University Institutional Review Board (protocol \irbnumber), all participants gave written informed consent, and recordings are stored under pseudonymous codes on access-restricted storage.

\section*{Data and Code}
The raw and processed recordings and their documentation are available in the Texas State University digital repository (\url{https://hdl.handle.net/10877/25342}; CC BY-NC-SA 4.0). The acquisition applications and their source code, the analysis code that regenerates every figure, table and number of this paper, and the independent reimplementation are available in the same repository (\url{https://hdl.handle.net/10877/25343}; PolyForm Noncommercial 1.0.0; copyright Texas State University).

\section*{Acknowledgments}
This study was funded by Texas State University. AI tools used during this work: we used Claude (Anthropic) to help write analysis and plotting code, to draft and edit text, to generate tables from the analysis outputs and to cross-check results. We used OpenAI Codex for an independent reimplementation of the analyses, sensitivity analyses and manuscript revision. The authors reviewed and edited all content and take full responsibility for it.

\bibliographystyle{plainnat}
\bibliography{refs_v3}

\end{document}


\title{Supplementary Material for ``Eye Tracking Signal Quality Evaluation: Meta Quest Pro vs HTC VIVE Pro Eye vs HTC VIVE Focus Vision''}
\renewcommand{\shorttitle}{Supplementary Material: Eye Tracking Signal Quality Evaluation}
\date{October 2026}
\author{
  Aleksandr Chernysh \\
  Texas State University\\
  San Marcos, TX, USA \\
  \texttt{wll46@txstate.edu}
  \And
  Mehedi Hasan Raju \\
  Texas State University\\
  San Marcos, TX, USA \\
  \texttt{m.raju@txstate.edu}
  \And
  Kamrul Hasan \\
  Texas State University\\
  San Marcos, TX, USA \\
  \texttt{kamrul.hasan@txstate.edu}
  \AND
  Md Arif Shahriar \\
  Texas State University\\
  San Marcos, TX, USA \\
  \texttt{arif.shahriar@txstate.edu}
  \And
  Proloy Karmakar \\
  Texas State University\\
  San Marcos, TX, USA \\
  \texttt{proloy@txstate.edu}
  \And
  Isayas B. Adhanom \\
  Texas State University\\
  San Marcos, TX, USA \\
  \texttt{isayas@txstate.edu}
  \AND
  Oleg Komogortsev \\
  Texas State University\\
  San Marcos, TX, USA \\
  \texttt{ok@txstate.edu}
}
\maketitle

\noindent This supplement gives the details that the main paper refers to. As in the main paper, QP, FV and PE denote the Meta Quest Pro, HTC VIVE Focus Vision and HTC VIVE Pro Eye, RAN127 and RAN63 the random-saccade blocks on the gray and the dark background, and U$|$E the user and error percentiles. Every table and figure is generated from the results of this study; Section~\ref{s:checks} lists the sources. Participants are identified only by pseudonymous codes.

\section{Protocol, Apparatus and Geometry}
\label{s:protocol}

\paragraph{Literature search} On 14 Oct 2026 we searched the web, arXiv, the ACM Digital Library, IEEE Xplore, SpringerLink, MDPI, OSF and PsyArXiv for evaluations of the eye tracker of the HTC VIVE Focus Vision. Queries combined ``Focus Vision'' or ``VIVE Focus Vision'' with ``eye tracking'', ``accuracy'', ``precision'', ``data quality'' and ``validation''. We found no study that reports its accuracy or precision; one preprint used the headset for gaze input without reporting signal quality. We found no source stating whether the FV tracker and the VIVE Focus~3 eye-tracker add-on share hardware.

{\sloppy\paragraph{Apparatus} Table~\ref{st:hardware} lists the installed software. The QP logger read gaze through the OVRPlugin call \texttt{GetEyeGazesState}. The versions are facts about the recorded builds; we did not test other versions.\par}

\begin{table}[!htbp]
\caption{Installed eye-tracking software and sampling of the three headsets (RAN127, medians over 30 recordings). Interface versions and application builds are those of the recorded builds. Timestamp cadence: reciprocal of the median interval between distinct device timestamps on QP and FV; on PE, whose clock has a resolution of 1\,ms (intervals of 8 or 9\,ms, Table~\ref{st:timing}), the increase of the sample sequence number per second over spans of 120 samples. The effective rate counts unique usable samples per second of application time.}
\label{st:hardware}
\centering\footnotesize
\begin{tabular}{@{}lp{0.13\textwidth}p{0.25\textwidth}p{0.14\textwidth}lrr@{}}
\toprule
Headset & Platform & Eye-tracking interface & Device clock & App build & Cadence (Hz) & Effective (Hz) \\
\midrule
QP & standalone (Android) & OVRPlugin (C\# wrapper 1.79.0), Unity Oculus XR Plugin 3.2.2 & gaze-sample time (s) & 2.10.4 & 90 & 89 \\
FV & standalone (Android) & HTC VIVE OpenXR 2.5.1 eye-tracker extension & pupil-sample time (ns) & 3.8.4 & 122 & 114 \\
PE & PC-tethered (Windows, SteamVR) & SRanipal SDK 1.3.2.0, runtime 1.3.6.12 & SRanipal tracker time (ms) & 1.9.4 & 120 & 119 \\
\bottomrule
\end{tabular}
\end{table}

\paragraph{Geometry} Targets lie on a sphere of radius 1\,m around the head origin. The cyclopean eye origin reported by each interface lies at a different depth relative to the head origin, so the eye-to-target distance differs between headsets (Table~\ref{st:geometry}). Angular error is measured from the eye origin of each device; ring assignment and eccentricity use the head origin.

\begin{table}[!htbp]
\caption{Geometry of the reported cyclopean eye origin, RAN127 (medians over 30 participants per headset): depth of the eye origin along the forward axis relative to the head origin (mm; positive: in front), distance from the eye origin to the target (m; targets lie on a 1\,m sphere around the head origin) and inter-pupillary distance derived from the logged eye origins (mm).}
\label{st:geometry}
\centering\small
\begin{tabular}{@{}lrrr@{}}
\toprule
Headset & Eye-origin depth (mm) & Eye--target distance (m) & IPD (mm) \\
\midrule
QP & 0 & 1.000 & 65 \\
FV & 23 & 0.978 & 66 \\
PE & $-36$ & 1.034 & 65 \\
\bottomrule
\end{tabular}
\end{table}

\paragraph{Headset order} Each of the six headset orders was used by five participants, so each headset was used ten times in each position. To separate position from headset, Table~\ref{st:order} centers each participant mean error on the median of its headset and compares the three positions within participants; Table~\ref{st:orderhl} gives the paired position differences with their intervals. No position contrast reached $\pholm<.05$.

\begin{table}[!htbp]
\caption{Participant mean error by the position of the headset in the session (first, second, third; 30 participants, each headset 10 times in each position), centered on the headset: participant mean error minus the median of that headset (deg; medians over participants), and the Friedman test across the three positions within participants (analysis A10). Paired contrasts are in Table~\ref{st:orderhl}.}
\label{st:order}
\centering\small
\begin{tabular}{@{}lrrrrr@{}}
\toprule
Block & First & Second & Third & $\chi^2$ & $p$ \\
\midrule
RAN127 & $+0.15$ & $-0.01$ & $-0.12$ & 4.20 & .122 \\
RAN63 & $+0.21$ & $-0.12$ & $-0.06$ & 0.47 & .792 \\
\bottomrule
\end{tabular}
\end{table}

\begin{table}[!htbp]
\caption{Paired differences between headset positions of the centered participant mean error (deg; 30 participants; analysis A10): Hodges--Lehmann estimate with exact 95\,\% interval, exact signed-rank $p$ and Holm-corrected $p$ over the three contrasts of a block.}
\label{st:orderhl}
\centering\small
\begin{tabular}{@{}llrrr@{}}
\toprule
Block & Positions & HL [exact 95\,\% CI] & $p$ & $\pholm$ \\
\midrule
RAN127 & second $-$ first & $-0.27$ [$-0.63$, $-0.04$] & .040 & .121 \\
 & third $-$ second & $+0.09$ [$-0.14$, $+0.35$] & .382 & .382 \\
 & third $-$ first & $-0.29$ [$-0.65$, $+0.09$] & .140 & .281 \\
RAN63 & second $-$ first & $-0.10$ [$-0.64$, $+0.42$] & .777 & 1.000 \\
 & third $-$ second & $-0.04$ [$-0.44$, $+0.41$] & .839 & 1.000 \\
 & third $-$ first & $-0.24$ [$-0.68$, $+0.23$] & .339 & 1.000 \\
\bottomrule
\end{tabular}
\end{table}

\paragraph{Validation gate} The first validation of every run was a gate (application code, identical on all headsets): it passed with a mean error of at most 3\dg{} and at least 30 usable samples and 80\,\% usable samples per target. The application allowed at most three gate attempts per run. In practice, after one or two failed attempts the experimenter either repeated the validation or recalibrated the tracker and restarted the run; recalibrations were not logged (the HTC applications record only an acknowledgement of the calibration instruction before each gate attempt). Later validations had no error threshold and were repeated only when data were insufficient. Table~\ref{st:gate} counts the attempts and failures. No run reached the three-attempt limit. On FV, four runs in three sessions were restarted after one or two failed attempts, and the session continued in a new run whose gate passed. On PE every failed attempt was followed by a pass in the same run. QP had 40 gate attempts in 30 sessions without a failure because every run begins with a new gate and 7 QP sessions were recorded in 2--4 runs (10 additional runs); the QP logs do not record why a run was restarted. Five task recordings were repeated with complete attempts (P003 FV and PE TEX, P012 QP BROW, P030 FV and PE BROW; the latest complete attempt is analyzed), and three attempts were aborted and left out (P001 QP RAN63, P003 and P005 QP RAN127).

\begin{table}[!htbp]
\caption{Five-point validations per headset over 30 sessions (analysis A11). Gate: first validation of a run (pass: mean error at most 3\dg{} and at least 30 usable samples and 80\,\% per target); failed gate attempts by reason, and sessions with at least one failed gate attempt. Later validations have no error threshold; the last two columns count those above 3\dg{} or with insufficient data. Gate attempts exceed 30 because restarted runs begin with a new gate.}
\label{st:gate}
\centering\footnotesize
\begin{tabular}{@{}lrrrrrrr@{}}
\toprule
Headset & Validations & Gate attempts & Gate $>$3\dg & Gate insufficient & Sessions with failed gate & Later $>$3\dg & Later insufficient \\
\midrule
QP & 152 & 40 & 0 & 0 & 0 & 4 & 1 \\
FV & 161 & 37 & 6 & 0 & 3 & 19 & 1 \\
PE & 184 & 40 & 1 & 8 & 6 & 3 & 22 \\
\bottomrule
\end{tabular}
\end{table}

\FloatBarrier
\section{Timing}
\label{s:timing}

Table~\ref{st:timing} summarizes the device clocks. The FV logger polls the interface twice per rendered frame and writes a row only when the pupil timestamp changes (Table~\ref{st:fvpoll}). Its 122\,Hz cadence is therefore the cadence of distinct pupil timestamps; whether every pupil timestamp carries a new gaze estimate cannot be verified from the log.

\begin{table}[!htbp]
\caption{Timing on the device clock. Median interval (ms) and its standard deviation (temporal precision; $\mu$s) between unique device timestamps in RAN127 (medians over 30 recordings), the same without intervals longer than 1.5 periods, and the number of intervals longer than 0.1\,s over all 150 recordings of a headset. The PE clock has a resolution of 1\,ms, so its intervals are 8 or 9\,ms; the median of 8\,ms therefore does not equal the reciprocal of its 120\,Hz cadence (Table~\ref{st:hardware}), and its SD without gaps reflects this resolution.}
\label{st:timing}
\centering\small
\begin{tabular}{@{}lrrrr@{}}
\toprule
Headset & Median interval (ms) & SD ($\mu$s) & SD without gaps ($\mu$s) & Intervals $>0.1$\,s \\
\midrule
QP & 11 & 2438 & 30 & 233 \\
FV & 8 & 1809 & 240 & 1 \\
PE & 8 & 471 & 471 & 0 \\
\bottomrule
\end{tabular}
\end{table}

\begin{table}[!htbp]
\caption{FV logger and pupil timestamps: medians over all 150 FV recordings. The logger polls the interface twice per rendered frame and writes a row only when the pupil timestamp changes.}
\label{st:fvpoll}
\centering\small
\begin{tabular}{@{}lr@{}}
\toprule
Measure & FV \\
\midrule
Interface polls per second & 180 \\
Rendered frames per second & 90 \\
Rows written per second & 114 \\
Intervals of one tick (6--10\,ms), \% & 94.4 \\
Intervals of two ticks ($\geq$14\,ms), \% & 5.5 \\
Pupil-timestamp cadence (Hz) & 122 \\
\bottomrule
\end{tabular}
\end{table}

\FloatBarrier
\section{Spatial Accuracy}
\label{s:accuracy}

Table~\ref{st:hl} gives the Hodges--Lehmann estimates of the headset contrasts reported in the main paper. Table~\ref{st:coverage} repeats the contrasts after leaving out windows with less than 80\,\% of the expected samples; only the QP--PE contrast in RAN127 crosses the 0.05 threshold. Table~\ref{st:coverageblocks} does the same for the differences of the block changes; there the PE--FV contrast lies on both sides of the threshold.

\begin{table}[!htbp]
\caption{Hodges--Lehmann (HL) estimates of the paired differences of participant mean error (deg) between headsets, 30 participants, with exact 95\,\% intervals from inverting the signed-rank test (analysis A3), rank-biserial $r$ and Holm-corrected exact signed-rank $p$. Negative values: the first headset has the smaller error.}
\label{st:hl}
\centering\small
\begin{tabular}{@{}llrrr@{}}
\toprule
Block & Pair & HL [exact 95\,\% CI] & $r$ & $\pholm$ \\
\midrule
RAN127 & QP$-$FV & $-1.09$ [$-1.43$, $-0.81$] & $-0.99$ & $<$.001 \\
 & QP$-$PE & $-0.28$ [$-0.59$, $+0.00$] & $-0.41$ & .052 \\
 & PE$-$FV & $-0.83$ [$-1.18$, $-0.51$] & $-0.88$ & $<$.001 \\
RAN63 & QP$-$FV & $-1.28$ [$-1.77$, $-0.77$] & $-0.75$ & $<$.001 \\
 & QP$-$PE & $-0.94$ [$-1.36$, $-0.50$] & $-0.71$ & $<$.001 \\
 & PE$-$FV & $-0.37$ [$-0.89$, $+0.10$] & $-0.32$ & .124 \\
\bottomrule
\end{tabular}
\end{table}

\begin{table}[!htbp]
\caption{Holm-corrected exact signed-rank $p$ of the headset contrasts of participant mean error when windows with less than 80\,\% of the expected samples are left out (expected number from the nominal rate or from the cadence of the recording). Source: independent reproduction (no results table covers this rule).}
\label{st:coverage}
\centering\small
\begin{tabular}{@{}p{0.36\textwidth}lrrr@{}}
\toprule
Window rule & Block & QP$-$FV & QP$-$PE & PE$-$FV \\
\midrule
All windows with data (main) & RAN127 & $<$.001 & .052 & $<$.001 \\
 & RAN63 & $<$.001 & $<$.001 & .124 \\
Windows with $\geq$80\,\% of nominal samples & RAN127 & $<$.001 & .047 & $<$.001 \\
 & RAN63 & $<$.001 & .001 & .114 \\
Windows with $\geq$80\,\% of cadence samples & RAN127 & $<$.001 & .047 & $<$.001 \\
 & RAN63 & $<$.001 & .001 & .124 \\
\bottomrule
\end{tabular}
\end{table}

\begin{table}[!htbp]
\caption{Holm-corrected exact signed-rank $p$ of the differences between headsets of the block change (RAN63 $-$ RAN127 participant mean error), with all windows and without windows that have less than 80\,\% of the expected samples (30 participants). Source: independent reproduction (no results table covers this rule).}
\label{st:coverageblocks}
\centering\small
\begin{tabular}{@{}p{0.42\textwidth}rrr@{}}
\toprule
Window rule & QP $-$ FV & QP $-$ PE & PE $-$ FV \\
\midrule
All windows with data (main) & .792 & .002 & .024 \\
Windows with $\geq$80\,\% of nominal samples & .855 & .005 & .047 \\
Windows with $\geq$80\,\% of cadence samples & .839 & .005 & .059 \\
\bottomrule
\end{tabular}
\end{table}

\subsection{User and Error Percentiles}
\label{s:ue}

Tables~\ref{st:ueRAN127} and~\ref{st:ueRAN63} give the full grid of user and error percentiles under the definition of the main paper \cite{aziz2024etra}, with bootstrap intervals, and under the rank definition of the preprint \cite{aziz2024arxiv}. The rank definition reports the whole error distribution of one participant and changes in steps when another participant moves to that rank.

\begin{table}[!htbp]
\caption{User and error percentiles of the window errors in RAN127 (deg; 30 participants per headset). First value: definition of the main paper (percentile across participants of each participant's E-th percentile; 95\,\% participant bootstrap interval, 10,000 resamples). After the slash: rank definition of the preprint (whole error distribution of the participant at rank 15, 23 or 29 of 30 by mean error).}
\label{st:ueRAN127}
\centering\footnotesize
\begin{tabular}{@{}rrlll@{}}
\toprule
$U$ & $E$ & QP & FV & PE \\
\midrule
50 & 50 & 0.83 [0.71, 1.00] / 0.81 & 2.10 [1.87, 2.38] / 2.12 & 1.07 [0.89, 1.20] / 1.15 \\
50 & 75 & 1.21 [1.02, 1.46] / 1.21 & 3.00 [2.75, 3.44] / 3.16 & 1.63 [1.51, 2.20] / 1.65 \\
50 & 95 & 2.21 [1.63, 2.84] / 2.83 & 4.41 [4.02, 5.51] / 4.49 & 4.04 [3.15, 4.68] / 3.57 \\
75 & 50 & 1.10 [0.89, 1.56] / 1.60 & 2.49 [2.28, 3.30] / 2.30 & 1.33 [1.15, 1.66] / 1.41 \\
75 & 75 & 1.75 [1.30, 2.51] / 2.07 & 3.52 [3.14, 4.47] / 3.47 & 2.37 [1.87, 3.10] / 2.70 \\
75 & 95 & 3.67 [2.39, 6.14] / 2.80 & 5.69 [4.83, 6.65] / 5.69 & 4.84 [4.27, 5.78] / 5.19 \\
95 & 50 & 1.86 [1.17, 2.00] / 1.92 & 3.36 [2.62, 3.53] / 3.35 & 2.19 [1.41, 3.17] / 1.83 \\
95 & 75 & 2.58 [2.25, 4.17] / 2.64 & 4.75 [3.97, 5.98] / 4.77 & 3.61 [2.61, 4.28] / 2.61 \\
95 & 95 & 6.71 [5.05, 7.56] / 7.56 & 8.37 [5.92, 14.53] / 8.70 & 8.02 [5.24, 12.48] / 9.06 \\
\bottomrule
\end{tabular}
\end{table}

\begin{table}[!htbp]
\caption{User and error percentiles of the window errors in RAN63 (deg; 30 participants per headset). First value: definition of the main paper (percentile across participants of each participant's E-th percentile; 95\,\% participant bootstrap interval, 10,000 resamples). After the slash: rank definition of the preprint (whole error distribution of the participant at rank 15, 23 or 29 of 30 by mean error).}
\label{st:ueRAN63}
\centering\footnotesize
\begin{tabular}{@{}rrlll@{}}
\toprule
$U$ & $E$ & QP & FV & PE \\
\midrule
50 & 50 & 0.87 [0.70, 1.05] / 1.02 & 2.07 [1.76, 2.48] / 1.63 & 1.78 [1.45, 2.16] / 1.66 \\
50 & 75 & 1.19 [1.06, 1.50] / 1.36 & 3.08 [2.62, 3.81] / 3.17 & 2.79 [2.27, 3.45] / 3.33 \\
50 & 95 & 1.98 [1.76, 3.41] / 1.93 & 4.56 [4.18, 5.05] / 4.12 & 5.27 [4.67, 6.01] / 5.03 \\
75 & 50 & 1.21 [0.94, 1.71] / 1.51 & 2.88 [2.35, 3.29] / 3.19 & 2.37 [1.94, 2.79] / 1.87 \\
75 & 75 & 1.93 [1.25, 2.82] / 1.94 & 4.14 [3.21, 5.05] / 3.94 & 3.57 [3.27, 4.16] / 3.10 \\
75 & 95 & 3.88 [2.80, 8.61] / 2.45 & 6.20 [4.75, 8.87] / 5.19 & 6.22 [5.63, 8.54] / 6.15 \\
95 & 50 & 2.50 [1.55, 6.05] / 2.88 & 3.92 [3.11, 4.60] / 4.60 & 3.19 [2.67, 3.52] / 2.78 \\
95 & 75 & 4.19 [2.26, 15.33] / 4.93 & 5.77 [4.32, 6.64] / 5.48 & 4.73 [3.80, 5.85] / 5.85 \\
95 & 95 & 10.54 [5.40, 29.52] / 7.99 & 10.52 [7.25, 13.96] / 8.16 & 10.26 [6.51, 11.10] / 10.11 \\
\bottomrule
\end{tabular}
\end{table}

\subsection{Off-Target Windows}
\label{s:otf}

A window is flagged when its error exceeds the median error of its spatial neighbors in the same recording by more than 5\dg{} (four neighbors for targets inside the outer ring, the two ring neighbors for outer targets). Table~\ref{st:otf} shows how many windows were flagged and how the medians and \UE{95}{95} change without them.

\begin{table}[!htbp]
\caption{Off-target windows: windows flagged by the neighbor rule (gaze away from the target), and the median participant mean error and \UE{95}{95} (deg) with all windows and without the flagged windows (30 participants per headset). The flagged windows are kept in the main analysis.}
\label{st:otf}
\centering\small
\begin{tabular}{@{}llrrrrr@{}}
\toprule
Block & Headset & Flagged windows & Median, all & Median, without & \UE{95}{95}, all & \UE{95}{95}, without \\
\midrule
RAN127 & QP & 13 & 1.04 & 1.04 & 6.71 & 6.44 \\
 & FV & 4 & 2.41 & 2.41 & 8.37 & 8.28 \\
 & PE & 6 & 1.50 & 1.49 & 8.02 & 7.15 \\
RAN63 & QP & 10 & 1.07 & 0.97 & 10.54 & 10.64 \\
 & FV & 2 & 2.26 & 2.25 & 10.52 & 10.52 \\
 & PE & 9 & 2.28 & 2.23 & 10.26 & 10.28 \\
\bottomrule
\end{tabular}
\end{table}

\FloatBarrier
\section{Accuracy Across the Field of View}
\label{s:fov}

Figure~\ref{sf:ecc} shows the error at each position against eccentricity with the random-intercept fits of Table~\ref{st:ecc}. Table~\ref{st:radius} gives the coverage radius for every ring. Figure~\ref{sf:maps} shows the RAN63 maps, and Figure~\ref{sf:rankmaps} the maps of single participants.

\suppfig{fig_ecc.pdf}{sf:ecc}{Window error against target eccentricity in RAN127 and RAN63: median over 30 participants of the participant mean error (deg) at each of the 49 positions, and the random-intercept fits of Table~\ref{st:ecc}.}

\begin{table}[!htbp]
\caption{Random-intercept models of window error on target eccentricity (maximum likelihood, one model per headset and block, about 1,500 windows of 30 participants each): slope (deg per 10\dg) with Wald 95\,\% interval, intercept (deg) and standard deviation of the participant intercepts (deg). The model assumes one slope for all participants; the main paper reports the mean of participant slopes, which agrees.}
\label{st:ecc}
\centering\small
\begin{tabular}{@{}llrrr@{}}
\toprule
Block & Headset & Slope [95\,\% CI] & Intercept & Participant SD \\
\midrule
RAN127 & QP & 0.35 [0.24, 0.46] & 0.69 & 0.55 \\
 & FV & 0.72 [0.58, 0.86] & 1.30 & 0.83 \\
 & PE & 0.81 [0.68, 0.94] & 0.26 & 0.64 \\
RAN63 & QP & 0.55 [0.37, 0.73] & 0.71 & 1.69 \\
 & FV & 0.82 [0.67, 0.98] & 1.36 & 1.01 \\
 & PE & 0.88 [0.74, 1.02] & 0.87 & 0.80 \\
\bottomrule
\end{tabular}
\end{table}

\begin{table}[!htbp]
\caption{Smallest radius (deg) around the target that contains 50, 90 and 95\,\% of all RAN127 window errors of a ring, windows pooled over 30 participants per headset. These radii describe windows, not users.}
\label{st:radius}
\centering\small
\begin{tabular}{@{}lrrrrrrrrr@{}}
\toprule
Ring & \multicolumn{3}{c}{QP} & \multicolumn{3}{c}{FV} & \multicolumn{3}{c}{PE} \\ & 50 & 90 & 95 & 50 & 90 & 95 & 50 & 90 & 95 \\
\midrule
center & 0.70 & 1.38 & 2.27 & 1.51 & 2.87 & 3.37 & 0.56 & 1.29 & 1.49 \\
inner & 0.76 & 1.61 & 1.93 & 1.53 & 3.42 & 4.64 & 0.73 & 1.72 & 2.27 \\
middle & 0.77 & 2.03 & 3.00 & 1.85 & 3.74 & 4.38 & 0.90 & 2.46 & 3.16 \\
outer & 1.00 & 2.81 & 4.71 & 2.61 & 5.01 & 6.11 & 1.45 & 4.43 & 5.73 \\
all & 0.87 & 2.35 & 3.61 & 2.09 & 4.42 & 5.48 & 1.04 & 3.39 & 4.67 \\
\bottomrule
\end{tabular}
\end{table}

\suppfig{fig_ran63_maps.pdf}{sf:maps}{Accuracy at each of the 49 target positions in RAN63: median over 30 participants of the participant mean error (deg), on the color scale of the RAN127 maps in the main paper (Figure~3). The two center presentations are averaged within the participant first.}

\suppfig[0.8\textwidth]{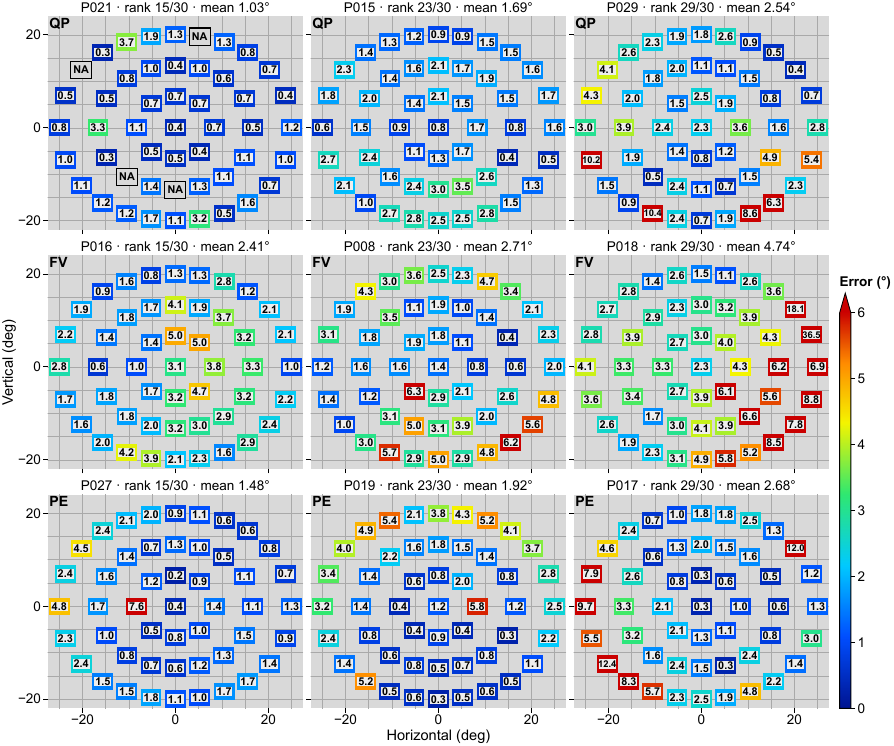}{sf:rankmaps}{Accuracy maps (deg, RAN127) of single participants at ranks 15, 23 and 29 of 30 by participant mean error, one row per headset; the panel title gives the participant code and mean error. Gray tiles marked NA have no usable window.}

\FloatBarrier
\section{Linearity, Offsets and Recalibration}
\label{s:linearity}

Table~\ref{st:linearity} gives the linearity slopes under both geometric conventions, and Table~\ref{st:r2} the goodness of the linear fits. With head-origin angles the horizontal slopes no longer differ between headsets in RAN127, while the vertical compression of FV remains. Table~\ref{st:centroid} shows that the convention moves two contrasts of centroid error across the significance threshold. Table~\ref{st:offsets} and Figure~\ref{sf:offsets} give the signed offsets, and Table~\ref{st:recal} the recalibration.

\begin{table}[!htbp]
\caption{Linearity: mean over 30 participants of the slope of the window gaze centroid against target position (ideal 1), with target angles from the device eye origin / from the head origin, and the Friedman $p$ across headsets under each convention. Head-origin values: re-analysis of the window table of the independent reproduction (matlab/eye\_origin\_sensitivity.m).}
\label{st:linearity}
\centering\footnotesize
\begin{tabular}{@{}llllll@{}}
\toprule
Block & Axis & QP & FV & PE & Friedman $p$ \\
\midrule
RAN127 & H & 0.965 / 0.965 & 0.942 / 0.962 & 1.009 / 0.969 & $<$.001 / .301 \\
 & V & 0.979 / 0.979 & 0.903 / 0.923 & 1.014 / 0.975 & $<$.001 / .001 \\
RAN63 & H & 0.948 / 0.949 & 0.926 / 0.945 & 1.028 / 0.988 & $<$.001 / .002 \\
 & V & 0.957 / 0.957 & 0.918 / 0.939 & 1.061 / 1.021 & $<$.001 / $<$.001 \\
\bottomrule
\end{tabular}
\end{table}

\begin{table}[!htbp]
\caption{Linearity fits: adjusted $R^2$ of the regression of the 50 window centroids on target position (horizontal H and vertical V head-local angles, target angles from the device eye origin), median over 30 participants (lowest participant in parentheses).}
\label{st:r2}
\centering\small
\begin{tabular}{@{}lllll@{}}
\toprule
Block & Axis & QP & FV & PE \\
\midrule
RAN127 & H & 0.998 (0.970) & 0.992 (0.827) & 0.994 (0.933) \\
 & V & 0.997 (0.878) & 0.988 (0.889) & 0.994 (0.897) \\
RAN63 & H & 0.998 (0.609) & 0.991 (0.837) & 0.994 (0.922) \\
 & V & 0.996 (0.251) & 0.985 (0.921) & 0.990 (0.932) \\
\bottomrule
\end{tabular}
\end{table}

\begin{table}[!htbp]
\caption{Effect of the geometric convention on the headset contrasts of centroid error (error of the window gaze centroid; 30 participants): Holm-corrected exact signed-rank $p$ with target angles from the eye origin and from the head origin. The two comparisons near the threshold cross it.}
\label{st:centroid}
\centering\small
\begin{tabular}{@{}llrr@{}}
\toprule
Block & Contrast & Eye origin & Head origin \\
\midrule
RAN127 & QP$-$FV & $<$.001 & $<$.001 \\
 & QP$-$PE & .040 & .073 \\
 & PE$-$FV & $<$.001 & $<$.001 \\
RAN63 & QP$-$FV & $<$.001 & $<$.001 \\
 & QP$-$PE & $<$.001 & .004 \\
 & PE$-$FV & .084 & .050 \\
\bottomrule
\end{tabular}
\end{table}

\begin{table}[!htbp]
\caption{Signed offsets (gaze minus target, head-local angles; deg): medians over 30 participants of the participant mean horizontal (H) and vertical (V) offset, signed-rank $p$ of V against zero, and the number of participants with gaze below the target. The PE $p$ is the same in both blocks because the signed-rank sums are mirrored (282 and 183).}
\label{st:offsets}
\centering\small
\begin{tabular}{@{}llrrrr@{}}
\toprule
Block & Headset & H & V & $p$ (V) & Below target \\
\midrule
RAN127 & QP & $-0.02$ & $-0.41$ & $<$.001 & 25 \\
 & FV & $+0.17$ & $+0.36$ & .009 & 10 \\
 & PE & $-0.03$ & $+0.01$ & .318 & 15 \\
RAN63 & QP & $-0.06$ & $-0.28$ & .026 & 21 \\
 & FV & $+0.13$ & $+1.15$ & $<$.001 & 3 \\
 & PE & $-0.07$ & $-0.36$ & .318 & 16 \\
\bottomrule
\end{tabular}
\end{table}

\suppfig{fig_offsets.pdf}{sf:offsets}{Systematic offset (gaze minus target, head-local angles) at each of the 49 positions, median over 30 participants of the participant mean, for RAN127 (top) and RAN63 (bottom); arrows are magnified four times (scale bar: 1\dg).}

\begin{table}[!htbp]
\caption{Linear user-supplied recalibration ($h'=a_1h+b_1v+c_1$, $v'=a_2h+b_2v+c_2$): median participant mean error (deg) before and after correction and the number of 30 participants whose error got worse. The in-sample fit evaluates on its own training data and is an upper bound, not a test. Paired changes (HL) are in Table~\ref{st:hlall}.}
\label{st:recal}
\centering\small
\begin{tabular}{@{}llrrr@{}}
\toprule
Variant & Headset & Before & After & Worse \\
\midrule
Spatial holdout (RAN127) & QP & 1.04 & 0.80 & 8 \\
 & FV & 2.41 & 1.48 & 0 \\
 & PE & 1.50 & 1.25 & 5 \\
Block transfer (RAN63) & QP & 1.07 & 1.00 & 11 \\
 & FV & 2.26 & 1.85 & 3 \\
 & PE & 2.28 & 1.67 & 3 \\
In-sample fit (RAN127) & QP & 1.04 & 0.77 & 4 \\
 & FV & 2.41 & 1.38 & 0 \\
 & PE & 1.50 & 1.15 & 3 \\
\bottomrule
\end{tabular}
\end{table}

\FloatBarrier
\section{Difference Between the Two Blocks}
\label{s:blocks}

RAN63 always followed RAN127 and the reading task. Table~\ref{st:blocks} compares accuracy, precision and loss between the blocks, and Table~\ref{st:rings} the change by ring. Table~\ref{st:hlall} gives the change between the first and second half of each block, and Table~\ref{st:p005} repeats the block comparison without the participant who was recalibrated on FV between the blocks. Table~\ref{st:pupil} gives the pupil diameter.

\begin{table}[!htbp]
\caption{RAN127 against RAN63 (30 participants): medians of the participant values in each block, the number of participants with a higher value in RAN63 and the exact signed-rank $p$ (uncorrected). HL estimates of the accuracy change are in Table~\ref{st:hlall}.}
\label{st:blocks}
\centering\small
\begin{tabular}{@{}llrrrr@{}}
\toprule
Measure & Headset & RAN127 & RAN63 & Higher in RAN63 & $p$ \\
\midrule
Mean error (deg) & QP & 1.04 & 1.07 & 16 & .360 \\
 & FV & 2.41 & 2.26 & 18 & .109 \\
 & PE & 1.50 & 2.28 & 26 & $<$.001 \\
RMS (deg) & QP & 0.26 & 0.24 & 14 & .655 \\
 & FV & 0.04 & 0.05 & 19 & .119 \\
 & PE & 0.17 & 0.23 & 24 & $<$.001 \\
STD (deg) & QP & 0.48 & 0.43 & 18 & .839 \\
 & FV & 0.28 & 0.31 & 17 & .349 \\
 & PE & 0.50 & 0.60 & 22 & .031 \\
Samples lost (\%) & QP & 1.3 & 2.7 & 21 & .092 \\
 & FV & 6.0 & 6.4 & 25 & $<$.001 \\
 & PE & 1.0 & 2.0 & 20 & .070 \\
\bottomrule
\end{tabular}
\end{table}

\begin{table}[!htbp]
\caption{Participant mean error by ring (deg; 30 participants): medians in RAN127 and RAN63 and the Hodges--Lehmann estimate of the paired change RAN63 $-$ RAN127 with exact 95\,\% interval; $p$: exact signed-rank test, Holm-corrected over the four rings of a headset. Ring eccentricity: center 0\dg, inner 6.7--8.3\dg, middle 13.3--16.7\dg, outer 20.0--25.0\dg.}
\label{st:rings}
\centering\small
\begin{tabular}{@{}llrrrr@{}}
\toprule
Headset & Ring & RAN127 & RAN63 & Change, HL [95\,\% CI] & $\pholm$ \\
\midrule
QP & center & 0.71 & 0.63 & $-0.03$ [$-0.19$, $+0.17$] & 1.000 \\
 & inner & 0.78 & 0.87 & $+0.05$ [$-0.11$, $+0.25$] & 1.000 \\
 & middle & 0.91 & 0.86 & $+0.09$ [$-0.08$, $+0.27$] & 1.000 \\
 & outer & 1.18 & 1.21 & $+0.09$ [$-0.07$, $+0.30$] & 1.000 \\
FV & center & 1.52 & 1.38 & $-0.01$ [$-0.51$, $+0.33$] & 1.000 \\
 & inner & 1.69 & 1.77 & $+0.10$ [$-0.23$, $+0.43$] & 1.000 \\
 & middle & 2.00 & 2.14 & $+0.23$ [$-0.00$, $+0.54$] & .220 \\
 & outer & 2.86 & 2.90 & $+0.13$ [$-0.08$, $+0.35$] & .813 \\
PE & center & 0.55 & 1.03 & $+0.49$ [$+0.22$, $+0.84$] & $<$.001 \\
 & inner & 0.75 & 1.38 & $+0.62$ [$+0.41$, $+0.91$] & $<$.001 \\
 & middle & 1.10 & 1.79 & $+0.67$ [$+0.44$, $+0.92$] & $<$.001 \\
 & outer & 2.05 & 2.98 & $+0.76$ [$+0.52$, $+1.06$] & $<$.001 \\
\bottomrule
\end{tabular}
\end{table}

\begin{table}[!htbp]
\caption{Hodges--Lehmann estimates with exact 95\,\% intervals for all other paired differences in the main paper (analysis A3b; deg, slopes in deg per 10\dg). Changes are second minus first block, corrected minus uncorrected error, raised minus relaxed eyebrows, and second minus first half of a block (presentations 25--49 against 0--24); differences of changes compare two headsets within participants. $p$: exact signed-rank test, Holm-corrected within each family of three.}
\label{st:hlall}
\centering\footnotesize
\begin{tabular}{@{}p{0.30\textwidth}lrrr@{}}
\toprule
Family & Contrast & $N$ & HL [exact 95\,\% CI] & $\pholm$ \\
\midrule
Block change (RAN63 $-$ RAN127) & QP & 30 & $+0.10$ [$-0.06$, $+0.27$] & .360 \\
 & FV & 30 & $+0.14$ [$-0.04$, $+0.38$] & .219 \\
 & PE & 30 & $+0.70$ [$+0.47$, $+0.98$] & $<$.001 \\
Difference of block changes & PE $-$ QP & 30 & $+0.56$ [$+0.31$, $+0.86$] & .002 \\
 & PE $-$ FV & 30 & $+0.48$ [$+0.12$, $+0.88$] & .024 \\
 & FV $-$ QP & 30 & $+0.03$ [$-0.20$, $+0.30$] & .792 \\
No outer ring, RAN127 & QP $-$ FV & 30 & $-0.83$ [$-1.18$, $-0.63$] & $<$.001 \\
 & QP $-$ PE & 30 & $-0.06$ [$-0.27$, $+0.22$] & .670 \\
 & PE $-$ FV & 30 & $-0.86$ [$-1.17$, $-0.53$] & $<$.001 \\
No outer ring, RAN63 & QP $-$ FV & 30 & $-1.02$ [$-1.39$, $-0.71$] & $<$.001 \\
 & QP $-$ PE & 30 & $-0.67$ [$-1.00$, $-0.29$] & .006 \\
 & PE $-$ FV & 30 & $-0.37$ [$-0.93$, $+0.05$] & .100 \\
Eccentricity slope (per 10\dg) & FV $-$ QP & 30 & $+0.39$ [$+0.18$, $+0.60$] & .004 \\
 & PE $-$ QP & 30 & $+0.44$ [$+0.21$, $+0.68$] & .001 \\
 & PE $-$ FV & 30 & $+0.10$ [$-0.20$, $+0.37$] & .529 \\
Recalibration, spatial holdout & QP & 30 & $-0.15$ [$-0.23$, $-0.07$] & .003 \\
 & FV & 30 & $-0.82$ [$-1.06$, $-0.65$] & $<$.001 \\
 & PE & 30 & $-0.29$ [$-0.41$, $-0.17$] & $<$.001 \\
Recalibration, block transfer & QP & 30 & $-0.04$ [$-0.10$, $+0.02$] & .191 \\
 & FV & 30 & $-0.41$ [$-0.64$, $-0.25$] & $<$.001 \\
 & PE & 30 & $-0.46$ [$-0.63$, $-0.30$] & $<$.001 \\
Eyebrow raise (raised $-$ relaxed) & QP & 30 & $+0.11$ [$-0.14$, $+0.66$] & .404 \\
 & FV & 30 & $+0.78$ [$+0.30$, $+1.58$] & .013 \\
 & PE & 29 & $+1.24$ [$+0.44$, $+6.12$] & .002 \\
Difference of eyebrow effects & FV $-$ QP & 30 & $+0.31$ [$-0.71$, $+1.06$] & .530 \\
 & PE $-$ QP & 29 & $+1.06$ [$+0.18$, $+4.36$] & .061 \\
 & PE $-$ FV & 29 & $+0.69$ [$-0.51$, $+4.85$] & .530 \\
Second $-$ first half, RAN127 & QP & 30 & $+0.02$ [$-0.08$, $+0.09$] & 1.000 \\
 & FV & 30 & $+0.01$ [$-0.14$, $+0.28$] & 1.000 \\
 & PE & 30 & $+0.15$ [$+0.01$, $+0.30$] & .109 \\
Second $-$ first half, RAN63 & QP & 30 & $-0.06$ [$-0.19$, $+0.06$] & .926 \\
 & FV & 30 & $+0.08$ [$-0.10$, $+0.29$] & .926 \\
 & PE & 30 & $-0.02$ [$-0.31$, $+0.19$] & .926 \\
\bottomrule
\end{tabular}
\end{table}

\begin{table}[!htbp]
\caption{Block comparison without the participant who was recalibrated on FV between the blocks (P005): median participant mean error (deg) in each block, participants with higher error in RAN63, and Holm-corrected signed-rank $p$.}
\label{st:p005}
\centering\small
\begin{tabular}{@{}lrrrrr@{}}
\toprule
Headset & $N$ & RAN127 & RAN63 & Higher in RAN63 & $\pholm$ \\
\midrule
QP & 29 & 1.06 & 1.10 & 15 & .358 \\
FV & 29 & 2.42 & 2.19 & 17 & .351 \\
PE & 29 & 1.52 & 2.23 & 25 & $<$.001 \\
\bottomrule
\end{tabular}
\end{table}

\begin{table}[!htbp]
\caption{Pupil diameter (mean of left and right over usable window samples; medians over 30 participants) in the two blocks, participants with a larger pupil in RAN63 and the exact signed-rank test (analysis A6). PE reports millimeters; the FV interface documents millimeters, but the scale was not validated, so FV values are compared only within FV. QP logs no pupil size.}
\label{st:pupil}
\centering\small
\begin{tabular}{@{}lrrrrr@{}}
\toprule
Headset & RAN127 & RAN63 & Larger in RAN63 & $W$ & $p$ \\
\midrule
FV & 3.86 & 5.17 & 29 & 1 & $<$.001 \\
PE & 3.61 & 4.71 & 30 & 0 & $<$.001 \\
\bottomrule
\end{tabular}
\end{table}

\paragraph{Validation before each task} Table~\ref{st:valid} gives the error of the five-point validation shown before each task and Table~\ref{st:validhl} its change between successive validations. On PE the validation error did not change from before RAN127 to before TEX, rose for the validation before RAN63, the only PE validation on the dark background, and fell again before BROW, later in the session but on gray. On QP, whose validation before RAN63 was also on the dark background, the change before RAN63 was $+0.08\dg$ [$-0.03$, $+0.22$] ($\pholm=.474$). RAN63 is always third, so background and position in the session are confounded; the cause of the change is not tested.

\begin{table}[!htbp]
\caption{Error of the five-point validation shown before each task (deg; mean over the five targets, 0.5--1.5\,s after the first sample at each target): median [IQR] over 30 participants per headset (analysis A9). Background: every validation was shown on the background of the following task (dark only before RAN63). QP values are recomputed from the task files, FV and PE values are those logged by the application.}
\label{st:valid}
\centering\footnotesize
\begin{tabular}{@{}lllll@{}}
\toprule
Before & Background & QP & FV & PE \\
\midrule
RAN127 & gray & 0.84 [0.57, 1.08] & 1.84 [1.50, 2.26] & 0.97 [0.79, 1.17] \\
TEX & gray & 0.71 [0.63, 1.20] & 1.82 [1.52, 2.48] & 1.02 [0.73, 1.23] \\
RAN63 & dark & 0.82 [0.63, 1.23] & 2.02 [1.45, 2.78] & 1.35 [1.05, 1.78] \\
BROW & gray & 0.88 [0.69, 1.09] & 1.96 [1.55, 2.42] & 1.11 [0.71, 1.58] \\
WINK & gray & 1.03 [0.64, 1.33] & 2.04 [1.65, 2.75] & 1.13 [0.79, 1.57] \\
\bottomrule
\end{tabular}
\end{table}

\begin{table}[!htbp]
\caption{Change of the validation error between successive validations (later minus earlier; deg; 30 participants per headset; analysis A9): Hodges--Lehmann estimate with exact 95\,\% interval, and $\pholm$ (exact signed-rank test, Holm-corrected over the four steps of a headset). Rows name the task that follows each validation.}
\label{st:validhl}
\centering\footnotesize
\begin{tabular}{@{}llll@{}}
\toprule
Step & QP & FV & PE \\
\midrule
TEX $-$ RAN127 & $+0.06$ [$-0.09$, $+0.27$], .503 & $+0.14$ [$-0.07$, $+0.36$], .686 & $+0.00$ [$-0.09$, $+0.09$], .984 \\
RAN63 $-$ TEX & $+0.08$ [$-0.03$, $+0.22$], .474 & $+0.07$ [$-0.13$, $+0.40$], 1.000 & $+0.35$ [$+0.14$, $+0.55$], .003 \\
BROW $-$ RAN63 & $-0.05$ [$-0.19$, $+0.05$], .474 & $-0.08$ [$-0.36$, $+0.18$], 1.000 & $-0.23$ [$-0.40$, $-0.09$], .009 \\
WINK $-$ BROW & $+0.12$ [$-0.01$, $+0.21$], .402 & $+0.24$ [$+0.04$, $+0.52$], .042 & $+0.06$ [$-0.07$, $+0.27$], .880 \\
\bottomrule
\end{tabular}
\end{table}

\FloatBarrier
\section{Precision and Temporal Structure}
\label{s:precision}

\paragraph{Spectra} Following \citet{lohr2019}, we selected for each participant and headset the three most stable one-second segments of RAN127 (lowest radial interquartile range, all samples usable, no gap longer than 2.5 sample periods). We removed the mean and a second-order trend, applied a Hann window and computed single-sided magnitude spectra. The spectral ratio in Table~\ref{st:precision} interpolates every spectrum at the same integer frequencies (1--4\,Hz and 20--39\,Hz); this corrects a frequency-bin mismatch of the FV spectra in the original analysis. Figure~\ref{sf:spectra} shows the mean spectra on the same integer-frequency grid.

\begin{table}[!htbp]
\caption{Temporal structure in RAN127 (30 participants per headset): medians of the participant mean RMS and STD (deg), median [interquartile range] of the participant STD/RMS ratio over both blocks (white noise: 0.71), spectral ratio of the mean magnitude at 20--39\,Hz to that at 1--4\,Hz (90 segments per headset; source: independent reproduction, corrected frequency bands) and median [IQR] optimal shift (ms). Last column: participants (of 30) whose FV spectral ratio is below their ratio on QP or PE.}
\label{st:precision}
\centering\small
\begin{tabular}{@{}lrrlrlr@{}}
\toprule
Headset & RMS & STD & STD/RMS & Spectral ratio & Optimal shift & FV ratio lower \\
\midrule
QP & 0.26 & 0.48 & 1.3 [1.2, 1.6] & 0.281 & 220 [212, 233] & 30 \\
FV & 0.04 & 0.28 & 9.8 [8.5, 10.4] & 0.021 & 268 [257, 279] & -- \\
PE & 0.17 & 0.50 & 2.4 [2.2, 2.7] & 0.093 & 234 [217, 247] & 28 \\
\bottomrule
\end{tabular}
\end{table}

\suppfig[0.55\textwidth]{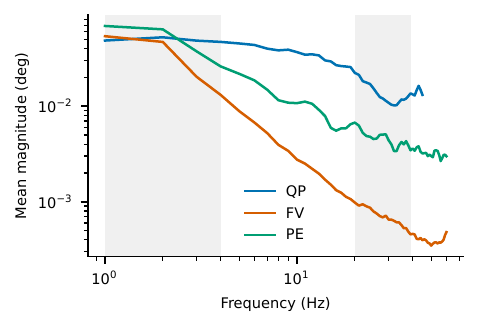}{sf:spectra}{Mean magnitude spectra of horizontal cyclopean gaze in stable one-second RAN127 segments (90 segments per headset), interpolated at integer frequencies as for the spectral ratio in Table~\ref{st:precision} (source: independent reproduction). Shading marks the two bands of the spectral ratio.}

\paragraph{Repeated values} Table~\ref{st:a4} shows that identical successive FV samples are rare and that removing them leaves the RMS unchanged. The longest run of identical values was 3 samples on FV and 2 on PE; QP had none.

\begin{table}[!htbp]
\caption{Repeated gaze values in the RAN windows (analysis A4; 30 participants per headset, both blocks): successive usable sample pairs, pairs with identical cyclopean directions and with a change below 0.001\dg{} (per 10,000 pairs; median [IQR] over participants), the longest run of identical values (samples), and the RAN127 RMS (deg) with all pairs and without identical pairs.}
\label{st:a4}
\centering\small
\begin{tabular}{@{}lrlrrrr@{}}
\toprule
Headset & Pairs & Identical & $<$0.001\dg & Longest run & RMS, all & RMS, non-identical \\
\midrule
QP & 127,513 & 0.0 [0.0, 0.0] & 29.4 & 1 & 0.26 & 0.26 \\
FV & 167,168 & 1.8 [1.8, 4.9] & 168.6 & 3 & 0.04 & 0.04 \\
PE & 170,513 & 0.0 [0.0, 0.0] & 31.5 & 2 & 0.17 & 0.17 \\
\bottomrule
\end{tabular}
\end{table}

\paragraph{Increments} Table~\ref{st:a12} describes successive changes of the gaze direction. FV increments were positively correlated over at least three samples in every participant, and QP and PE increments were not. Equal successive increments were 1.8\,\% on FV (QP 0.1\,\%, PE 0.4\,\%), far below the about 50\,\% expected under linear interpolation at half rate. The data therefore argue against linear interpolation, but they do not separate low-pass filtering from non-linear interpolation.

\begin{table}[!htbp]
\caption{Increments of gaze direction in the RAN127 windows (analysis A12; head-local horizontal and vertical angles, runs of at least 8 contiguous usable rows; medians over 30 participants per headset; descriptive). Equal increments: share of second differences below 0.0001\dg, about the float resolution of a unit vector; linear interpolation of a stream at half the rate would make about half of them zero. RMS ratio: RMS of second differences over RMS of first differences (white noise: 1.73). $r_k$: autocorrelation of the mean-removed increments at lag $k$ (mean over windows); last column: participants with $r_1>0$.}
\label{st:a12}
\centering\small
\begin{tabular}{@{}lrrlrrr@{}}
\toprule
Headset & Equal increments (\%) & RMS ratio & $r_1$ [IQR] & $r_2$ & $r_3$ & $r_1>0$ \\
\midrule
QP & 0.1 & 1.19 & $-0.08$ [$-0.10$, $-0.01$] & $-0.01$ & $-0.24$ & 6 \\
FV & 1.8 & 1.03 & $+0.29$ [$+0.24$, $+0.32$] & $+0.37$ & $+0.29$ & 30 \\
PE & 0.4 & 1.40 & $-0.06$ [$-0.10$, $-0.03$] & $-0.01$ & $+0.00$ & 5 \\
\bottomrule
\end{tabular}
\end{table}

\FloatBarrier
\section{Data Loss}
\label{s:loss}

The expected number of samples in a recording is the device-clock span times the timestamp cadence of the recording. Lost samples are invalid, frozen (missing slots in intervals that contain repeated timestamps) or dropped (the remaining gaps). Table~\ref{st:loss} gives the total loss with its upper quartile, Table~\ref{st:qpint} the QP interruptions and Table~\ref{st:windows} the short windows without valid data.

\begin{table}[!htbp]
\caption{Total data loss per task in percent of the expected samples on the device clock (invalid, frozen and dropped samples): median and 75th percentile over 30 participants. Categories and maxima are in Table~3 of the main paper.}
\label{st:loss}
\centering\small
\begin{tabular}{@{}lrrrrrr@{}}
\toprule
Task & \multicolumn{2}{c}{QP} & \multicolumn{2}{c}{FV} & \multicolumn{2}{c}{PE} \\ & Median & Q75 & Median & Q75 & Median & Q75 \\
\midrule
RAN127 & 1.3 & 2.2 & 6.0 & 6.5 & 1.0 & 2.0 \\
TEX & 0.5 & 1.1 & 6.0 & 6.7 & 1.3 & 1.9 \\
RAN63 & 2.7 & 4.2 & 6.4 & 7.4 & 2.0 & 3.3 \\
BROW & 0.5 & 3.6 & 7.3 & 8.6 & 1.9 & 4.8 \\
WINK & 2.8 & 4.0 & 30.6 & 31.8 & 27.4 & 29.9 \\
\bottomrule
\end{tabular}
\end{table}

\begin{table}[!htbp]
\caption{QP intervals longer than 0.1\,s between unique device timestamps over all 150 QP recordings (30 participants, five tasks), by composition: number, median and longest duration (s).}
\label{st:qpint}
\centering\small
\begin{tabular}{@{}lrrr@{}}
\toprule
Composition & Intervals & Median (s) & Longest (s) \\
\midrule
Repeated timestamp, then timestamp zero & 133 & 1.02 & 6.04 \\
Repeated timestamp only & 100 & 0.19 & 1.00 \\
\bottomrule
\end{tabular}
\end{table}

\begin{table}[!htbp]
\caption{Short windows without a valid sample (\% of windows; mean over the 120 recordings per headset of all tasks except WINK) for three window lengths.}
\label{st:windows}
\centering\small
\begin{tabular}{@{}lrrr@{}}
\toprule
Window & QP & FV & PE \\
\midrule
100\,ms & 1.2 & 0.4 & 2.5 \\
250\,ms & 1.0 & 0.3 & 1.2 \\
500\,ms & 0.7 & 0.2 & 0.9 \\
\bottomrule
\end{tabular}
\end{table}

\paragraph{Keeping repeated QP rows} Table~\ref{st:a1} recomputes the QP accuracy with the repeated-timestamp rows kept as samples. The RAN63 tail grows because 13 windows that were empty in the main analysis then contain only frozen gaze. Rows with timestamp zero carry no gaze geometry and no valid flag and cannot be added.

\begin{table}[!htbp]
\caption{QP accuracy (deg) with rows that repeat a device timestamp removed (main analysis) or kept as samples (analysis A1; 30 participants, 1,500 windows per block).}
\label{st:a1}
\centering\small
\begin{tabular}{@{}llrrrrr@{}}
\toprule
Block & Repeated rows & \UE{50}{50} & \UE{50}{95} & \UE{95}{95} & Median mean & Windows with data \\
\midrule
RAN127 & removed (main) & 0.83 & 2.21 & 6.71 & 1.04 & 1,491 \\
 & kept & 0.83 & 2.27 & 6.68 & 1.07 & 1,500 \\
RAN63 & removed (main) & 0.87 & 1.98 & 10.54 & 1.07 & 1,485 \\
 & kept & 0.87 & 2.23 & 12.52 & 1.11 & 1,498 \\
\bottomrule
\end{tabular}
\end{table}

\paragraph{Application frames} Table~\ref{st:a5} relates long application frames to the two-tick FV intervals and to the QP interruptions. The QP count of 230 covers intervals between fresh timestamps of the task rows; the main paper's count of 233 also includes 3 intervals that start at the first row of a task, whose timestamp had already appeared before the task.

\begin{table}[!htbp]
\caption{Application frames and data loss (analyses A5 and A5b; 150 recordings per headset). A long frame lasts more than 1.5 times the median frame of its recording. Poll pattern: the row is written at the first poll of a frame, the second poll of the frame returns the same sample, and the first poll of the next frame returns a sample two ticks later.}
\label{st:a5}
\centering\small
\begin{tabular}{@{}p{0.55\textwidth}rr@{}}
\toprule
Measure & FV & QP \\
\midrule
Median frame duration (ms) & 11 & 11 \\
Long frames per 10,000 frames & 2.3 & 2.0 \\
Two-tick FV intervals, \% of intervals (RAN127) & 5.1 & -- \\
Two-tick FV intervals in long frames, per 10,000 (RAN127) & 1.7 & -- \\
Skipped FV ticks in long frames, \% (RAN127) & 0.1 & -- \\
Two-tick FV intervals with the poll pattern, \% (RAN127) & 98.2 & -- \\
Gap between the two polls of a frame (ms) & 9 & -- \\
Gap to the first poll of the next frame (ms) & 2 & -- \\
QP interruptions in task rows / starting in a long frame & -- & 230 / 0 \\
\bottomrule
\end{tabular}
\end{table}

\FloatBarrier
\section{Eyebrow Raising and Eye Closure}
\label{s:facial}

Table~\ref{st:brow} compares the rule of the main paper (no exclusion) with the exclusion mask of the preprint \cite{aziz2024arxiv}. The mask removes the participant--headset records with the largest raised errors, which lowers the raised medians on all headsets. Table~\ref{st:wink} gives the eye-closure validity and counts, and Table~\ref{st:a2} the QP gaze during closures. Table~\ref{st:blink} shows that the QP validity flag also stayed set during the 5-s blink screens before every validation and task, and that the QP confidence did not mark the instructed closures. Meta's face-tracking eye-closure weights were not logged.

\begin{table}[!htbp]
\caption{Eyebrow raising with two participant rules: no exclusion (main paper) and the exclusion mask of the preprint (a participant--headset record is removed if any relaxed window before a raise or any raised window is missing or above 10\dg). Complete pairs, median error (deg) relaxed and raised, participants with higher raised error, and exact signed-rank $p$ (uncorrected).}
\label{st:brow}
\centering\small
\begin{tabular}{@{}llrrrrr@{}}
\toprule
Headset & Rule & Pairs & Relaxed & Raised & Raised higher & $p$ \\
\midrule
QP & none (main) & 30 & 0.96 & 1.07 & 18 & .404 \\
 & exclusion mask & 20 & 0.72 & 0.69 & 11 & .388 \\
FV & none (main) & 30 & 1.42 & 2.36 & 24 & .007 \\
 & exclusion mask & 20 & 1.32 & 1.90 & 16 & .011 \\
PE & none (main) & 29 & 0.60 & 2.51 & 24 & $<$.001 \\
 & exclusion mask & 12 & 0.56 & 0.95 & 9 & .110 \\
\bottomrule
\end{tabular}
\end{table}

\begin{table}[!htbp]
\caption{Eye closure (WINK; 30 participants, 5 closures each): share of rows flagged valid (medians over participants), closure onset (FV: first invalid row; PE: first row with both openness values below one half; not detectable on QP; this onset differs from the loss latency of the main paper, which is the first of three consecutive invalid rows and is 0.68\,s on PE), and closure counts (source of counts: independent reproduction).}
\label{st:wink}
\centering\small
\begin{tabular}{@{}lrrr@{}}
\toprule
Measure & QP & FV & PE \\
\midrule
Valid rows 0.5--1.5\,s after ``close'' (\%) & 100.0 & 41.8 & 16.6 \\
Valid rows 1.0--2.0\,s after ``close'' (\%) & 100.0 & 1.5 & 0.0 \\
Detected closure onset after the cue (s) & -- & 0.92 & 0.73 \\
Closures with recorded rows & 150 & 149 & 150 \\
Closures with sustained loss of validity & 0 & 148 & 150 \\
Closures that recovered & 0 & 148 & 150 \\
\bottomrule
\end{tabular}
\end{table}

\begin{table}[!htbp]
\caption{QP gaze with eyes open (last 0.5\,s before ``close'') and closed (1.0--2.0\,s after ``close''): medians [IQR] over 30 participants of the median over five closures, participants with a higher value when closed, and the signed-rank $p$ (analysis A2). All rows carried geometry and a valid flag.}
\label{st:a2}
\centering\small
\begin{tabular}{@{}lllrr@{}}
\toprule
Measure & Open & Closed & Closed higher & $p$ \\
\midrule
Sample-to-sample RMS (\dg) & 0.17 [0.16, 0.20] & 1.18 [0.99, 1.37] & 30 & $<$.001 \\
STD (\dg) & 0.23 [0.19, 0.28] & 2.98 [1.99, 3.55] & 30 & $<$.001 \\
Error to target (\dg) & 0.62 [0.50, 1.19] & 9.62 [7.73, 13.20] & 30 & $<$.001 \\
Distance from gaze before ``close'' (\dg) & 0.20 [0.17, 0.26] & 9.74 [7.36, 12.50] & 30 & $<$.001 \\
Linear trend speed (\dg/s) & 0.72 [0.50, 1.00] & 5.76 [3.72, 8.07] & 30 & $<$.001 \\
\bottomrule
\end{tabular}
\end{table}

\begin{table}[!htbp]
\caption{QP validity flag and gaze during the 5-s ``PLEASE BLINK'' screens before every validation and task (analysis A8; medians over screens or, for gaze, over participant medians), and the QP confidence during instructed eye closure (analysis A12; medians over 30 participants). The confidence takes only the values 0 and 0.998. FV and PE logged no gaze during the blink screens, and there is no independent record of the blinks on QP. The screens have no fixation target, so the gaze dispersion mixes free viewing and blinks.}
\label{st:blink}
\centering\small
\begin{tabular}{@{}p{0.62\textwidth}r@{}}
\toprule
Measure & QP \\
\midrule
Blink screens (2 per task, 5 tasks, 30 participants) & 300 \\
Duration per screen, median (s) & 4.99 \\
Rows flagged valid, median / lowest screen (\%) & 100.0 / 99.3 \\
Screens with $\geq$3 consecutive invalid rows & 2 \\
Gaze STD on the screen (\dg), median [IQR] & 5.90 [5.13, 6.89] \\
Samples $>$5\dg{} from the screen median direction (\%) & 32.4 \\
{}\quad of these below the median direction (\%) & 53.5 \\
Largest deviation from the screen median direction (\dg) & 18.19 \\
\addlinespace
WINK closed (1.0--2.0\,s after ``close''): rows with confidence 0.998 (\%) & 100.0 \\
WINK open (2--5\,s after ``open''): rows with confidence 0.998 (\%) & 100.0 \\
\bottomrule
\end{tabular}
\end{table}

\FloatBarrier
\section{Per-Eye Rays and Reliability}
\label{s:reliability}

Table~\ref{st:eyes} gives the angle between the left and right gaze rays. On QP it is nearly the same in every participant and within every window, so the two rays are not independent measurements. The table also gives the accuracy of each eye and the split-half reliability of participant mean error.

\begin{table}[!htbp]
\caption{Per-eye rays and reliability in RAN127 (30 participants per headset). Vergence: angle between the left and right gaze rays over usable window samples, median over participants [range of participant medians] and the median within-window interquartile range (deg; analysis A7). Accuracy: median participant mean error (deg) of the cyclopean, left and right signal on the same samples. Split-half reliability of participant mean error (spatially balanced halves of the positions, Spearman--Brown corrected).}
\label{st:eyes}
\centering\small
\begin{tabular}{@{}lrrlr@{}}
\toprule
Headset & Vergence [range] & Within-window IQR & Cyclopean / left / right & Split-half \\
\midrule
QP & 3.06 [3.05--3.07] & 0.10 & 1.04 / 1.06 / 1.07 & 0.85 \\
FV & 1.83 [1.61--2.30] & 0.17 & 2.41 / 2.48 / 2.64 & 0.95 \\
PE & 3.71 [1.74--4.69] & 1.08 & 1.50 / 1.84 / 1.60 & 0.93 \\
\bottomrule
\end{tabular}
\end{table}

\FloatBarrier
\section{Sources and Reproduction}
\label{s:checks}

\paragraph{Sources} All tables and figures are generated by the code released with the paper (one script per figure; \texttt{supplement\_tables/make\_supplement\_tables.py} for the tables; \texttt{MANIFEST.md} maps each figure and table to its script and inputs) from the results: the per-participant and per-window tables of the main analysis, the report tables built from the same per-sample data, and the additional analyses A1--A12 (stream cleaning, eye closure, Hodges--Lehmann estimates, repeated values, application frames, pupil, vergence, blink screens, validations before each task, headset position, validation gate and increments). Where no results table exists, the cell comes from the independent reproduction and the caption says so: the coverage rule (Tables~\ref{st:coverage} and~\ref{st:coverageblocks}), the corrected spectral ratios and spectra (Table~\ref{st:precision}, Figure~\ref{sf:spectra}), the eye-closure counts (Table~\ref{st:wink}) and the head-origin linearity and centroid contrasts (Tables~\ref{st:linearity} and~\ref{st:centroid}, re-analysis of the reproduction's window table). The table generator can list every cell with its source file, selector and rounding rule (option \texttt{-{}-manifest}); every cell was recomputed independently from that list before submission.

\paragraph{Independent reproduction} The second implementation was written by an AI coding tool (OpenAI Codex) from the same written specification, without importing the analysis code. From the raw recordings it recomputed the accuracy and user and error percentiles, precision, timing and data loss, the optimal shift, the off-target rule, linearity, the eccentricity models, recalibration and the eyebrow and eye-closure analyses. It reproduced the values of the main analysis and corrected three points: the frequency bands of the spectral ratio, the eye-closure counts and onset rule, and the definitions of the effective rate and the expected number of samples. It did not recompute the analyses A1--A12, the exact Hodges--Lehmann intervals, the split-half reliability or the per-eye accuracy; these come from the main analysis code only.

\bibliographystyle{plainnat}
\bibliography{refs_v3}